\documentclass[12pt]{article}
\usepackage{url}
\input{epsf}

\font\bm=cmmib10 at 10pt
\font\bms=cmmib10 at 7pt \textfont9=\bm \scriptfont9=\bms
\usepackage{amsfonts}
\usepackage{multirow}
\mathchardef\balpha= "790B
\mathchardef\bbeta= "790C
\mathchardef\bTheta= "7902
\mathchardef\bzeta= "7910
\mathchardef\bOmega= "790A
\mathchardef\bGamma= "7900
\mathchardef\bDelta= "7901
\mathchardef\bPhi= "7908
\mathchardef\bphi= "791E
\mathchardef\bomega= "7921
\mathchardef\bxi= "7918
\mathchardef\bet= "7911
\mathchardef\brho= "791A
\mathchardef\btau= "791C
\mathchardef\bmu= "7916
\mathchardef\bvarpi= "7924

\def \lvec{(\kern-.26em(}
\def\pmb#1{\setbox0=\hbox{#1}%
\def \lvec{(\kern-.26em(}
\kern-.025em\copy0\kern-\wd0
\kern.05em\copy0\kern-\wd0
\kern-.025em\raise.0433em\box0 }
\mathchardef\btheta= "7912
\usepackage{amsmath}
\usepackage{authblk}

\usepackage{graphicx}
\usepackage{dcolumn}

\begin{document}

\title{Relative Potency of Greenhouse Molecules}
\author[1]{W. A. van Wijngaarden}
\author[2]{W. Happer}
\affil[1]{Department of Physics and Astronomy, York University, Canada, wavw@yorku.ca}
\affil[2]{Department of Physics, Princeton University, USA, happer@Princeton.edu}
\renewcommand\Affilfont{\itshape\small}
\date{\today}
\maketitle

\noindent The forcings due to changing concentrations of Earth's five most important, naturally occurring greenhouse gases, H$_2$O, CO$_2$, O$_3$, N$_2$O and CH$_4$ as well as CF$_4$ and SF$_6$ were evaluated for the case of a cloud-free atmosphere.  The calculation used over 1.5 million lines having strengths as low as $10^{-27}$ cm.  For a hypothetical, optically thin atmosphere, where there is negligible saturation of the absorption bands, or interference of one type of greenhouse gas with others, the per-molecule forcings are of order $10^{-22}$ W for H$_2$O, CO$_2$, O$_3$, N$_2$O and CH$_4$ and of order $10^{-21}$ W for CF$_4$ and SF$_6$.  For current atmospheric concentrations, the per-molecule forcings of the abundant greenhouse gases H$_2$O and CO$_2$ are suppressed by four orders of magnitude.  The forcings of the less abundant greenhouse gases, O$_3$, N$_2$O and CH$_4$, are also suppressed, but much less so.  For CF$_4$ and SF$_6$, the suppression is less than an order of magnitude because the concentrations of these gases is very low.  For current concentrations, the per-molecule forcings are two to four orders of magnitude greater for O$_3$, N$_2$O, CH$_4$, CF$_4$ and SF$_6$ than those of H$_2$O or CO$_2$.  Doubling the current concentrations of CO$_2$, N$_2$O or CH$_4$ increases the forcings by a few per cent.  A concentration increase of either CF$_4$ or SF$_6$ by a factor of 100 yields a forcing nearly an order of magnitude smaller than that obtained by doubling CO$_2$.  Important insight was obtained using a harmonic oscillator model to estimate the power radiated per molecule.  Unlike the most intense bands of the 5 naturally occurring greenhouse gases, the frequency-integrated cross sections of CF$_4$ and SF$_6$ were found to noticeably depend on temperature.  
%
\newpage
\section{Introduction}
Accurate calculations of radiative forcing are essential to estimate future climate change \cite{IPCC,Hodnebrog}.  This paper examines the effect of changing greenhouse gas concentrations on thermal radiation for the case of a clear sky.  It considers the five most important naturally occurring greenhouse gases: H$_2$O, CO$_2$, O$_3$, N$_2$O and CH$_4$ as well as CF$_4$ and SF$_6$ that are nearly entirely of anthropogenic origin.  The atmospheric concentrations of these gases has been observed to be steadily increasing \cite{MaunaLoa, Worton}.  The concentrations of CF$_4$ and SF$_6$ are too low to significantly affect the present climate but have increased by over 25\% and 100\% respectively since 2000 .  These two molecules also have atmospheric lifetimes well in excess of 1,000 years \cite{Muhle, Kovacs}.  The Kyoto protocol seeks to limit emissions of the gases considered in this study \cite{IPCC}. 

Radiative forcings are strongly affected by saturation of the absorption bands and spectral overlap with other greenhouse gases.  Recently, this was found to significantly affect methane forcing \cite{Etminan2016}.  There have also been conflicting forcing estimates of the less abundant gases such as SF$_6$ \cite{Kovacs}. 
The spectra of greenhouse gases consists of hundreds of thousands of individual rovibrational spectral lines.  The most accurate forcings are found by performing line by line calculations that have been described by various authors \cite{Edwards1992,Collins2006,Schreier2014,vW2020}. 

This study downloaded the line strengths and transition frequencies of over 1.5 million rovibrational lines from the most recent HITRAN \cite{HITRAN} and VAMDC \cite{CNRS} databases to calculate the per-molecule forcings of the most important greenhouse gas molecules.  Each greenhouse gas concentration was varied from the optically thin limit where there is negligible saturation or interference of one type of greenhouse gas with others; to current levels.  The ``instantaneous" forcings resulting from doubling concentrations were compared to those published in the literature. 

This paper is organized as follows.  First, we describe the altitudinal profiles of the atmospheric temperature and the concentrations of the various greenhouse gases.  Next, the line intensities are briefly discussed.  The following section outlines how radiative forcing is determined.  The main part of the paper describes the 
concentration dependence of the radiative forcing.  Finally, the use of a harmonic oscillator model to accurately estimate the power radiated per greenhouse molecule without the need for detailed line intensity information is presented.

\begin{figure}[t]
\includegraphics[height=100mm,width=1\columnwidth]{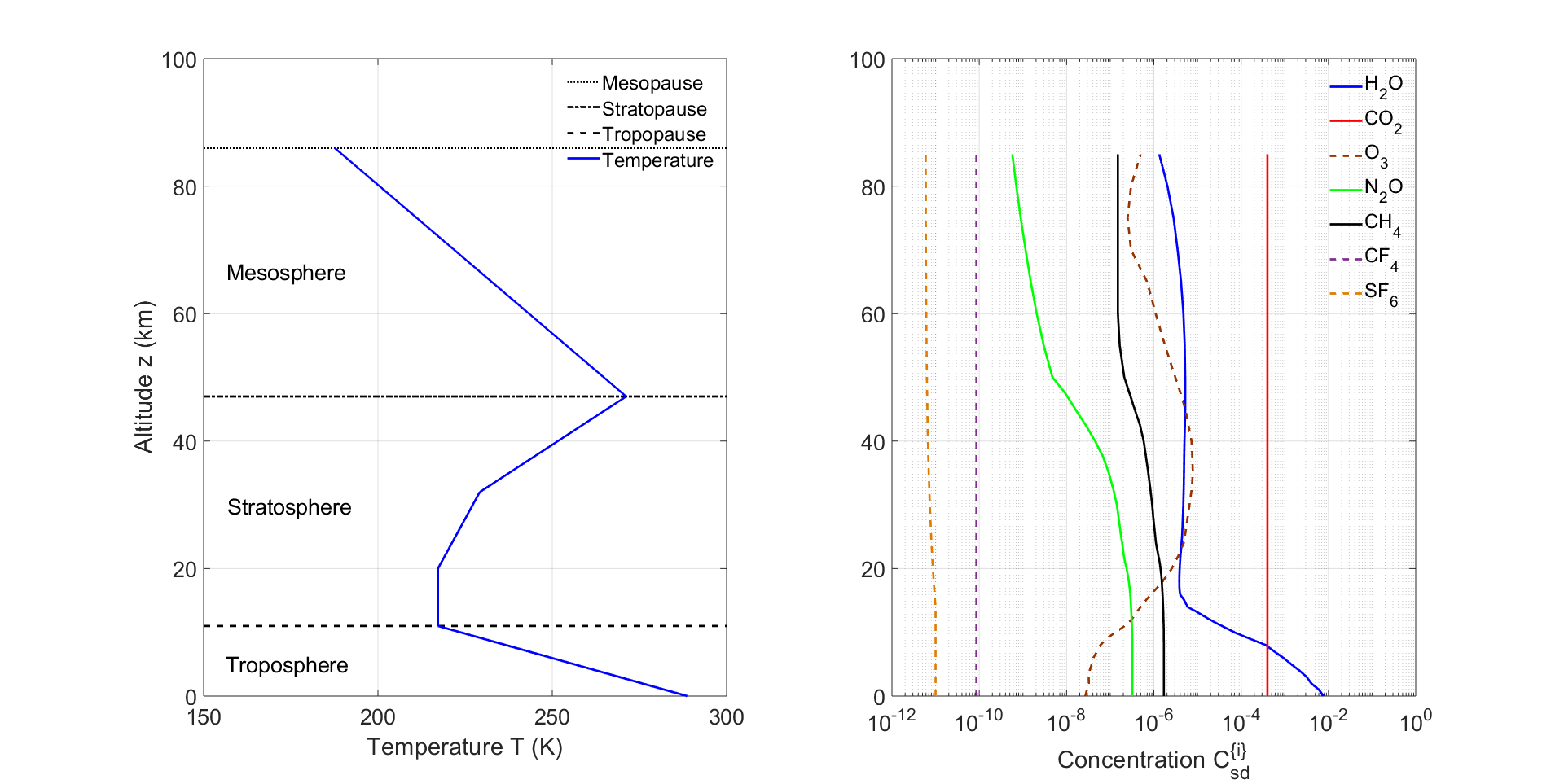}
\caption{{\bf Left.} A midlatitude atmospheric temperature profile, $T=T(z)$. The Earth's mean surface temperature  $T(0) = 288.7$ K.  {\bf Right.} Standard observed concentrations, $C^{\{i\}}_{\rm sd}$ for greenhouse molecules versus altitude $z$. 
\label{GGNT}}
\end{figure}

\section{Altitude Profiles of Temperature and Greenhouse Gases \label{he}}

Radiation transfer in the cloud-free atmosphere is controlled by the temperature $T=T(z)$ at the altitude $z$.  Fig. \ref{GGNT} shows the midlatitude atmospheric temperature profile \cite{Temp}.  The atmosphere was divided into 5 layers each having constant lapse rate.  We characterize the temperature profile with six breakpoints, with temperatures $\theta_{\alpha} =$ [288.7, 217.2, 217.2, 229.2, 271.2, 187.5] in units of Kelvins at altitudes ${\bf \zeta}_{\alpha}=[0, 11, 20, 32, 47, 86]$ in km, where $\alpha = 0, 1, ..., 5$.  Each of the 5 atmospheric layers was further subdivided into 100 sublayers.  

\begin{table}
\begin{center}
\begin{tabular}{|c|c| c|}
 \hline
 $i$&Molecule& $\hat N^{\{i\}}_{\rm sd}$ (molecules/cm$^2$)\\ [0.5ex]
 \hline\hline
 1& H$_2$O &$4.70\times 10^{22}$\\
 \hline
 2& CO$_2$ &$8.54\times 10^{21}$\\
 \hline
 3&O$_3$   &$9.18\times 10^{18}$\\
 \hline
 4&N$_2$O  &$6.56\times 10^{18}$\\
 \hline
 5&CH$_4$  &$3.73\times 10^{19}$\\
 \hline
 6&CF$_4$  &$1.84\times 10^{15}$\\
 \hline
 7&SF$_6$  &$2.10\times 10^{14}$\\
 \hline
 \end{tabular}
\end{center}
\caption{Column densities, $\hat N^{\{i\}}_{\rm sd}$, of the 5 most abundant greenhouse gases obtained using the standard altitudinal profiles of Fig. \ref{GGNT}.
\label{acd}}
\end{table}

The standard concentrations for the ith greenhouse gas, $C^{\{i\}}_{\rm sd}$, based on observations \cite{Kovacs,Anderson, Trudinger}, are shown as functions of altitude on the right of Fig. \ref{GGNT}.  The sea level concentrations for 2020 were estimated to be $7,750$ ppm of H$_2$O, $1.8$ ppm of CH$_4$, $0.32$ ppm of N$_2$O and $10$ ppt of SF$_6$. The O$_3$ concentration peaks at $7.8$ ppm at an altitude of 35 km, while the concentrations of CO$_2$ and CF$_4$ were $400$ ppm and $86$ ppt respectively, at all altitudes.  Integrating the concentrations over an atmospheric column having a cross sectional area of 1 cm$^2$ yields the column number density of the $i$th type of molecule $\hat N^{\{i\}}_{\rm sd}$ listed in Table \ref{acd}.

\section {Line Intensities}
 
\begin{figure}[t]
\includegraphics[height=90mm,width=1\columnwidth]{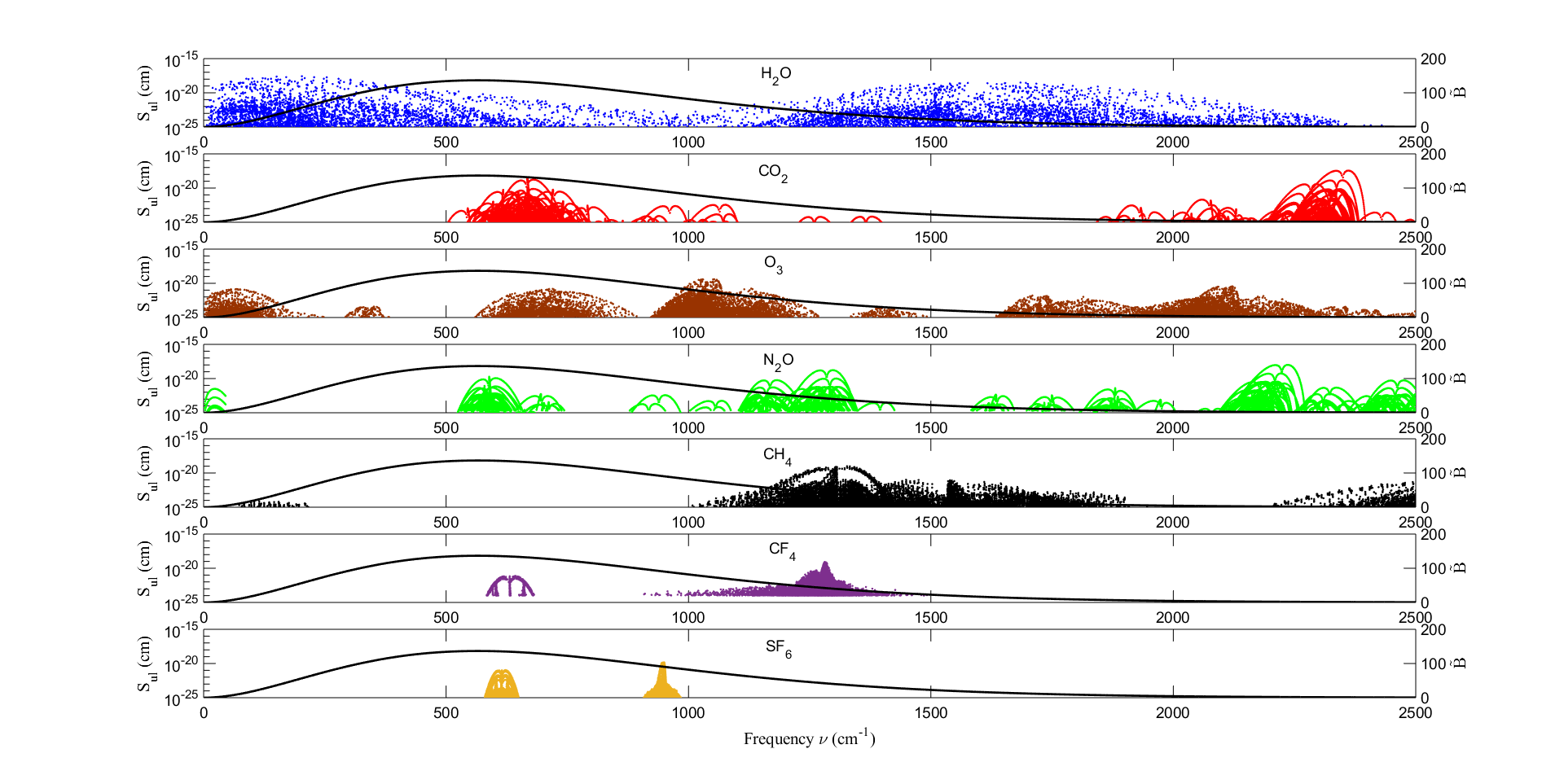}
\caption{Reference line intensities, $S_{ul}^{\{r\}}$ of (\ref{lbl36}) for H$_2$O, CO$_2$, O$_3$, N$_2$O, CH$_4$, CF$_4$ and SF$_6$ from the HITRAN data base \cite{HITRAN}.  The  horizontal coordinate of each point represents the Bohr frequency $\nu_{ul}$ of a transition from an upper level $u$ to a lower level $l$.  The vertical coordinate is the line intensity.  For CF$_4$, HITRAN only lists line intensities exceeding $10^{-24}$ cm.  For greater clarity we have plotted only 1/10, chosen at random, of the extremely large number of the O$_3$, CF$_4$ and SF$_6$ line intensities.  The numbers of lines (in parenthesis) used for this work were: H$_2$O (31,112), CO$_2$ (20,569), O$_3$ (210,295), N$_2$O (43,152), CH$_4$ (43,696) CF$_4$ (842,534) and SF$_6$ (307,780).  We also used SF$_6$ lines (336,027) from the VAMDC database which are nearly indistinugishable from those shown in this figure.  The smooth line is the Planck spectral intensity, $\tilde B$ of (\ref{b12}) in units of mW cm m$^{-2}$ sr$^{-1}$ for the reference temperature, $T^{\{r\}} = 296$ K.}
\label{Fig2}
\end{figure}

Fig. \ref{Fig2} illustrates the greenhouse gas lines considered in this work.  The  Bohr frequency $\nu_{ul}$ for a radiative transition from a lower level $l$ of energy $E_l$ to an upper level $u$ of energy $E_u$ of the same molecule is denoted by

\begin{equation}
\nu_{ul}=\frac{E_{ul}}{h c},\quad\hbox{where}\quad E_{ul}= E_u-E_l.
\label{lbl2}
\end{equation}

\noindent where the energy of a resonant photon is $E_{ul}$, $h$ is Planck's constant and $c$ is the speed of light.

The cross section, $\sigma^{\{i\}}=\sigma$, for the $i$th type of greenhouse molecule is written as the sum of partial cross sections $\sigma_{ul}$, corresponding to each Bohr frequency $\nu_{ul}$,

\begin{equation}
\sigma=\sum_{ul} \sigma_{ul}.
\label{lbl16}
\end{equation}

\noindent The partial cross section, $\sigma_{ul}$, is the product of a lineshape function, $G_{ul}=G_{ul}(\nu, \tau)$ where the optical depth $\tau$ is defined in the next section, and the
line intensity, $S_{ul}=S_{ul}(T)$,
\begin{equation}
\sigma_{ul}=G_{ul}S_{ul}.
\label{lbl18}
\end{equation}

\noindent The lineshape functions, $G_{ul}$, take into account the natural linewidth and Doppler Broadening as well as effects due to collisions \cite{vW2020}.  They are normalized to have unit area,

\begin{equation}
\int_0^{\infty}G_{ul}d\nu=1
\label{lbl20}
\end{equation}

\noindent and have units of cm.  The line intensity is

\begin{equation}
S_{ul} =\eta_{u}\pi r_e f_{ul} W_l\left(1-e^{-\nu_{ul}/\nu_T}\right)=\frac{\eta_u W_u\Gamma_{ul}E_{ul}}{4\pi \tilde B_{ul}}.
\label{lbl24}
\end{equation}

\noindent $S_{ul}$ has the units of cm and $r_e$ is the classical electron radius.  The isotopologue fractions are $\eta_u$.  For the most abundant isotopologues of CO$_2$,

\begin{equation}
\eta_{u}=\left \{\begin{array}{rr} 0.9843&\hbox{ for $^{16}$O $^{12}$C $^{16}$O}\\
0.0110 &\hbox{ for $^{16}$O $^{13}$C $^{16}$O}\\
0.0040 &\hbox{ for $^{16}$O $^{12}$C $^{18}$O}\\
0.0007 &\hbox{ for $^{16}$O $^{12}$C $^{17}$O}. \end{array}\right.
\label{lbl4}
\end{equation}

\noindent The last term of (\ref{lbl24}) contains the spectral Planck intensity evaluated at the frequency $\nu_{ul}$,

\begin{equation}
\tilde B_{ul}=\tilde B(\nu_{ul},T).
\label{lbl32}
\end{equation}

\noindent The Planck intensity is given by 
 
\begin{equation}
\tilde B(\nu,T)=\frac{2h c^2\nu^3}{e^{\nu c\, h/(k_{\rm B}T)}-1}
\label{b12}
\end{equation}

\noindent The radiation frequency, $\nu=1/\lambda$ is the inverse of the wavelength $\lambda$ and has units of cm$^{-1}$.  

\noindent The probability $W_n$ (with $n=u$ or $n=l$) to find a molecule in the rovibrational level $n$ is

\begin{equation}
W_n =\frac{g_n e^{-E_n/k_{\rm B}T}}{Q}.
\label{lbl26}
\end{equation}

\noindent Here $g_n$ is the statistical weight of the level $n$, the number of independent quantum states with the same energy $E_n$.
For molecules in the level $n$, the statistical weight can be taken to be
\begin{equation}
g_n=(2j_n+1)k_n,
\label{lbl27}
\end{equation}
where $j_n$ is the rotational angular momentum quantum number, and $k_n$ is the nuclear degeneracy factor, that depends on whether the spins of the nuclei are identical or not.  The partition function, $Q=Q(T)$, of the molecule is
\begin{equation}
Q=\sum_n g_n e^{-E_n/k_{\rm B}T}.
\label{lbl28}
\end{equation}

The oscillator strength, $f_{ul}$, of (\ref{lbl24}) is related to the matrix elements of the electric dipole moment {\bf M} of the molecule,  between the upper energy basis state $|j_u m_u\rangle$ with azimuthal quantum number $m_u$ and the lower energy basis state $|j_l m_l\rangle$, by
\begin{equation}
f_{ul}=\frac{4\pi\nu_{ul}}{3 g_lc\,r_e \hbar}
\sum_{m_u m_l}\langle u \,m_u|{\bf M}|l\, m_l\rangle\cdot\langle l\, m_l|{\bf M}|u\, m_u\rangle.
\label{lbl30}
\end{equation}
The quantum numbers $m_u$ label the various degenerate substates of the upper level $u$ and the $m_l$ label the substates of the lower level $l$.  If the levels are characterized by rotational quantum numbers $j_u$ and $j_l$, the quantum numbers $m_u$ and $m_l$ can be thought of as the corresponding azimuthal quantum numbers, for example, $m_u=j_u, j_u-1,\ldots,-j_u$.  The rate of spontaneous emission of photons when the molecule makes transitions from the upper level $u$ to the lower level $l$ is $\Gamma_{ul}$, the same as the Einstein $A$ coefficient. It is related to the oscillator strength by
\begin{equation}
\Gamma_{ul}=\frac{8\pi^2c\,r_e \nu_{ul}^2f_{ul}g_l}{ g_u}.
\label{lbl31}
\end{equation}

From inspection of (\ref{lbl24}) we see that the line intensity $S_{ul}=S_{ul}(T)$ at some arbitrary temperature $T$ is related to the intensity, $S_{ul}^{\{r\}}=S_{ul}(T^{\{r\}})$ at a reference temperature $T^{\{r\}}$ where the partition function of (\ref{lbl28}) is related to $Q^{\{r\}}=Q(T^{\{r\}})$  by
\begin{equation}
S_{ul} =S_{ul}^{\{r\}}\frac{Q^{\{r\}}}{Q}\left(\frac{e^{-E_l/k_BT}}{e^{-E_l/k_{\rm B} T^{\{r\}}}}\right)\left(\frac{1-e^{-\nu_{ul}/\nu_T}}{1-e^{-\nu_{ul}/\nu_{T^{\{r\}}}}}\right).
\label{lbl36}
\end{equation}

\noindent The HITRAN and VAMDC databases list line intensities at a reference temperature $T^{\{r\}}=296$ K.  This work considered all lines of the seven gases under consideration having intensities greater than $10^{-25}$ cm.  For H$_2$O, lines having intensities greater than $10^{-27}$ cm were included since water vapor has an order of magnitude greater density than any other greenhouse gas near the Earth's surface. 
  
\section{Calculation of Radiative Forcing}

Radiation transport is governed by the Schwarzschild equation \cite{Schwarzschild1906} in cloud-free air where scattering is negligible.

\begin{equation}
\cos \theta \frac{\partial \tilde I}{\partial \tau}=-(\tilde I-\tilde B)
\label{nsa4}
\end{equation}

\noindent Here ${\tilde I} = {\tilde I}(\nu,z,\theta)$ is the spectral intensity of a pencil of radiation of frequency between $\nu$ and $\nu + d\nu$ at altitude $z$.  The pencil makes an angle $\theta$ to the vertical.  The optical depth is defined by 

\begin{equation}
\tau(z,\nu)=\int_0^{z}dz' \kappa(z',\nu),
\label{b10}
\end{equation}

\noindent where the net attenuation coefficient due to molecules absorbing and reemitting light of frequency $\nu$ at altitude $z$ is given by 

\begin{equation}
\kappa(z,\nu)=\sum_i N^{\{i\}}(z)\sigma^{\{i\}}(z,\nu).
\label{b8}
\end{equation}

\noindent Here $N^{\{i\}}(z)$ is the density of greenhouse gas molecule of type $i$ and $\sigma^{\{i\}}=\sigma^{\{i\}}(z,\nu)$ is its absorption cross section for radiation of frequency $\nu$ at the altitude $z$ given by (\ref{lbl16}).  The cross section can depend strongly on altitude because temperature and pressure are functions of altitude.  Temperature controls the distribution of the molecules between translational, rotational and vibrational states.  Pressure, together with temperature, determines the width of the molecular resonance lines.  

The optical depth from the surface to the top of the radiative atmosphere, the altitude $z_{\rm mp}$ of the mesopause, is

\begin{equation}
\tau_{\infty}=\tau_{\rm mp}=\int_0^{z_{\rm mp}}dz'\kappa(z',\nu).
\label{b11}
\end{equation}

\noindent As indicated by the notation (\ref{b11}), we have assumed that the optical depth $\tau_{\rm mp}$ at the mesopause altitude $z_{\rm mp}$ differs negligibly from the optical depth $\tau_{\infty}$ at infinite altitude since there is so little opacity of the atmosphere above the mesopause.

The Schwarzschild equation (\ref{nsa4}) can be solved to find the intensity \cite{Buglia}

\begin{eqnarray}
\hbox{For $\varsigma>0$}:\quad\tilde I(\tau,\varsigma)&=& +\varsigma \int_0^{\tau}d\tau' e^{-\varsigma (\tau-\tau')}\tilde B(\tau')+e^{-\varsigma\tau}\tilde I(0,\varsigma)\label{vn48}\\
\hbox{For $\varsigma<0$}:\quad\tilde I(\tau,\varsigma)&=& -\varsigma \int_{\tau}^{\tau_{\infty}}d\tau' e^{-\varsigma (\tau-\tau')}\tilde B(\tau')\label{vn50}
\end{eqnarray}

\noindent where $\varsigma= \sec \theta$.  For simplicity, we assume the surface intensity is the product of $\tilde B_s=\tilde B(T_s)$, the Planck intensity (\ref{b12}) for a temperature $T_s$, and an angle independent emissivity $\epsilon_s=\epsilon_s(\nu)$,

\begin{equation}
\tilde I(0,\varsigma)= \epsilon_s\tilde B_s.
\label{vna51}
\end{equation}

\noindent Over most of the Earth's surface the thermal infrared emissivity $\epsilon_s$, is observed to be in the interval $[0.9 < \epsilon_s<1]$ \cite{Wilber}.  The emissivity $\epsilon_s$ was therefore set to 1.  We also assumed there is negligible temperature discontinuity between the surface and the air immediately above so that $\tilde B_s=\tilde B_0$.

The upwards flux defined by 

\begin{equation}
\tilde Z=\int_{4\pi}d\Omega\,\cos\theta\,\tilde I.
\label{b5}
\end{equation}

\noindent can be rewritten after substituting (\ref{vn48}) and (\ref{vn50}) into (\ref{vna51}) to give

\begin{eqnarray}
{{\tilde Z}\over {2 \pi}}&=&\int_0^{\tau}d\tau'E_2(\tau-\tau')\tilde B(\tau') - \int_{\tau}^{\tau_{\infty}} d\tau' E_2(\tau'-\tau)\tilde B(\tau')+ \epsilon_s\tilde B_sE_3(\tau)  \nonumber\\
&=&-\int_0^{\tau_{\infty}} d\tau' E_3(\vert \tau - \tau' \vert) {{\partial \tilde B(\tau')}\over{\partial \tau'}} +
\tilde B(\tau_{\infty}) E_3(\tau_{\infty} - \tau).
\label{vn14}
\end{eqnarray}

\noindent This equation is the fundamental expression for the net upward flux in an atmosphere with negligible scattering \cite{Buglia,Yoshikawa}.  The exponential-integral functions, $E_n(\tau)$, that account for slant paths of radiation between different altitudes are defined for integers $n=1,2,3,\ldots$ by
\begin{equation}
E_n(\tau)=\int_1^{\infty}d\varsigma\,\varsigma^{-n}\,e^{-\varsigma\tau}.
\label{b40}
\end{equation}

The spectral forcing, $\tilde F$, is defined as the difference between the spectral flux $\pi\tilde B_s$ through a transparent atmosphere from a black surface with temperature $T_s$, and the spectral flux $\tilde Z$ for an atmosphere with greenhouse gases,

\begin{equation}
\tilde F=\pi \tilde B_s-\tilde Z\\.
\label{vn58}
\end{equation}

\noindent The frequency integrals of the flux (\ref{b5}) and the forcing (\ref{vn58}) are

\begin{eqnarray}
Z&=&\int_0^{\infty}d\nu\,\tilde Z,\label{b58}\\
F&=&\int_0^{\infty}d\nu\,\tilde F=\sigma_{\rm SB}T_0^4-Z,\label{b60}\\
\end{eqnarray}

\noindent where $\sigma_{SB}$ is the Stefan Boltzmann constant.

\begin{figure}[t]
\includegraphics[height=100mm,width=1.0\columnwidth]{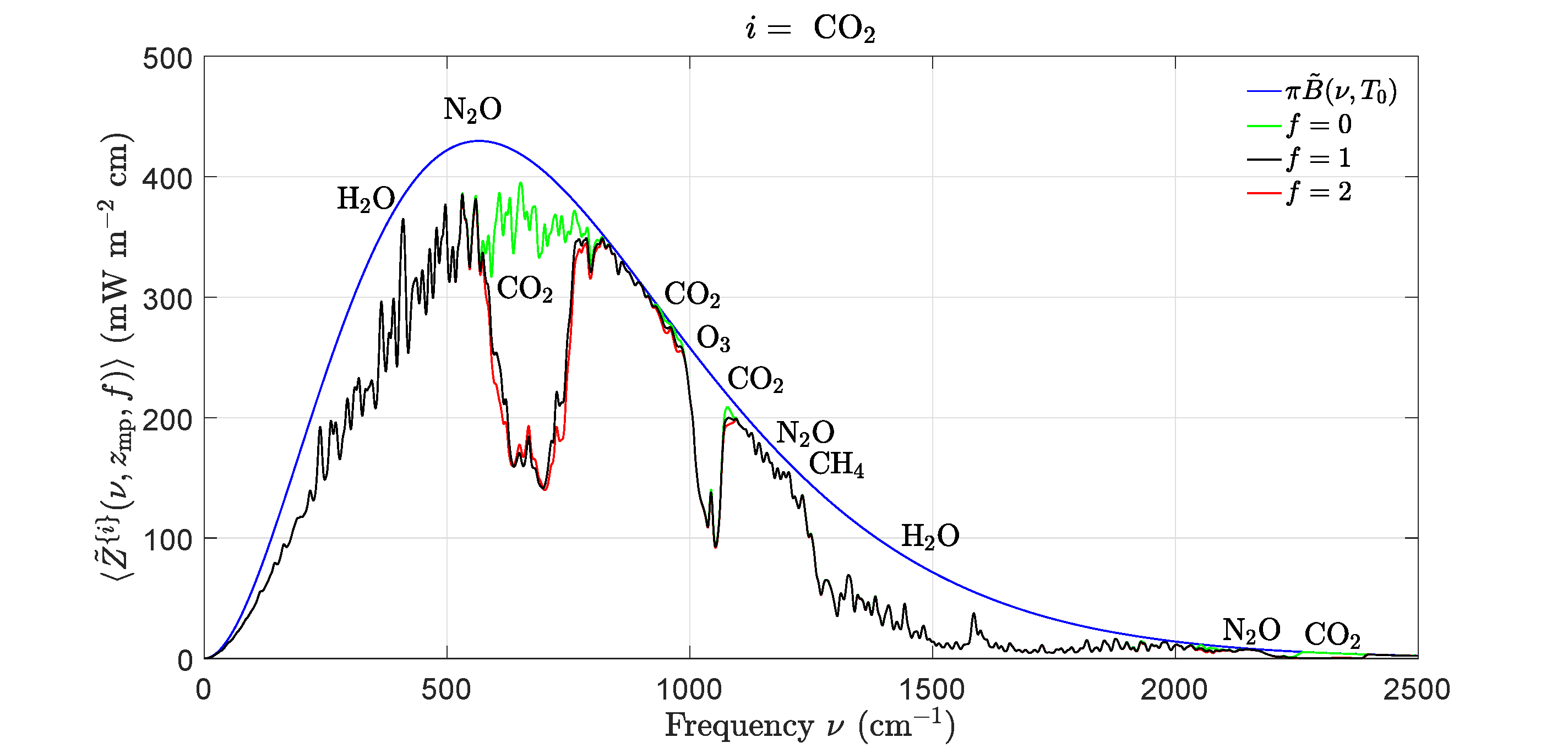} 
\caption{Effects of changing concentrations of carbon dioxide, CO$_2$ on the filtered spectral flux $\langle \tilde Z^{\{i\}}(\nu,z_{\rm mp},f)\rangle $ of (\ref{am10}) at the mesopause altitude, $z_{\rm mp}=86$ km.  The  width of the filter (\ref{am6}) was $\Delta\nu = 3$ cm$^{-1}$. The smooth blue line is the spectral flux, $\tilde Z =\pi\tilde B(\nu,T_0)$ from a surface at the temperature $T_0=288.7$ K for a transparent atmosphere with no greenhouse gases.  The green line is $\langle \tilde Z^{\{i\}}(\nu,z_{\rm mp},0)\rangle $ with the CO$_2$ removed but with all the other greenhouse gases at their standard concentrations.  The black line is
$\langle \tilde Z^{\{i\}}(\nu,z_{\rm mp},1)\rangle$ with all greenhouse gases at their standard concentrations. The red line is $\langle \tilde Z^{\{i\}}(\nu,z_{\rm mp},2)\rangle $ for twice the standard concentration of CO$_2$ but with all the other greenhouse gases at their standard concentrations.
Doubling the standard concentration of CO$_2$ (from 400 to 800 ppm) would cause a forcing increase (the area between the black and red lines)  of $\Delta F^{\{i\}} =2.97$ W m$^{-2}$, as shown in Table \ref{int}.
\label{CO2}}
\end{figure}

\begin{figure}[t] 
\includegraphics[height=100mm,width=1.0\columnwidth]{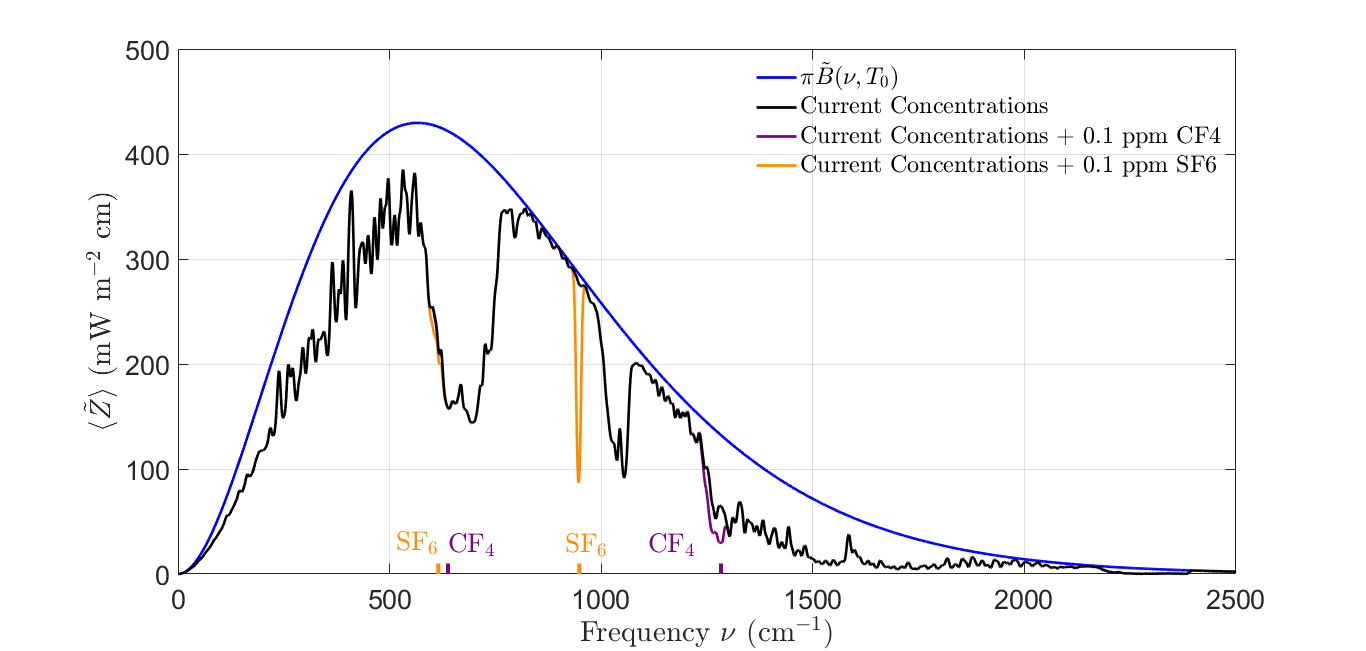}     
\caption{Effects of CF$_4$ and SF$_6$ on the filtered spectral flux $\langle \tilde Z^{\{i\}}(\nu,z_{\rm mp},f)\rangle $ of (\ref{am10}) at the mesopause altitude, $z_{\rm mp}=86$ km.  The blue and black lines have the same meanings as for Fig. \ref{CO2}.  The purple (orange) line is $\langle \tilde Z^{\{i\}}(\nu,z_{\rm mp},f)\rangle $ for a CF$_4$ (SF$_6$) concentration of 0.1 ppm which far exceeds the 2020 CF$_4$ (SF$_6$) concentration of 86 (10) ppt.  The purple and orange marks on the horizontal axis show the vibrational center frequencies of 615, 631, 948 and 1283 cm$^{-1}$.
\label{CF4SF6}}
\end{figure}

High resolution spectrometers seldom provide measurements with resolutions less than a few cm$^{-1}$.  It is therefore useful to plot filtered spectral quantities.

\begin{equation}	
\langle \tilde X\rangle(z,\nu) = \int_0^{\infty}d\nu\,'J(\nu,\nu\,')\tilde X(z,\nu\,').
\label{am2}
\end{equation}

\noindent The filter function $J(\nu,\nu\,')$ smooths out sharp changes with frequency.  It is normalized so that

\begin{equation}	
\int_{-\infty}^{\infty}d\nu J(\nu,\nu\,')=1.
\label{am4}
\end{equation}

\noindent From (\ref{am2}) and (\ref{am4}) we see that the unfiltered spectral flux $\tilde Z$ and filtered spectral flux $\langle \tilde Z\rangle$ have the same frequency integral

\begin{equation}	
Z=\int_{0}^{\infty}d\nu \tilde Z=\int_{0}^{\infty}d\nu \,\langle\tilde Z\rangle,
\label{am8}
\end{equation}

\noindent and represent the same total flux $Z$.  We found it convenient to use a Gaussian filter function,

\begin{equation}	
J(\nu,\nu\,')=\frac{e^{-(\nu-\nu\,')^2/2\Delta\nu^2}}{\sqrt{2\pi}\Delta\nu}
\label{am6}
\end{equation}

\noindent with a width parameter $\Delta \nu = 3$ cm$^{-1}$.

The effects on radiative transfer of changing the column density of the $i$th greenhouse gas to some multiple $f$ of the standard value, $\hat N^{\{i\}}_{\rm sd}$, can be displayed with filtered spectral fluxes

\begin{equation}
\langle\tilde Z^{\{i\}}(\nu,z,f)\rangle=\langle\tilde Z(\nu, z,\hat N^{\{1\}}_{\rm sd},\ldots,\hat N^{\{i-1\}}_{\rm sd},f\hat N^{\{i\}}_{\rm sd}, \hat N^{\{i+1\}}_{\rm sd},\ldots,
\hat N^{\{n\}}_{\rm sd})\rangle.
\label{am10}
\end{equation}

\noindent Fig. \ref{CO2} shows how varying the concentration of CO$_2$ affects the filtered spectral fluxes at the mesopause altitude, $z_{\rm mp} = 86$ km.  The present concentrations of CF$_4$ and SF$_6$ are too low to have any noticeable effect.  Fig. \ref{CF4SF6} shows the effect on the flux at the mesopause altitude if one were to increase the concentrations of CF$_4$ and SF$_6$ by nearly 3 and 4 orders of magnitude respectively, to 0.1 ppm. 

\begin{figure}[t]   
\includegraphics[height=100mm,width=1.0\columnwidth]{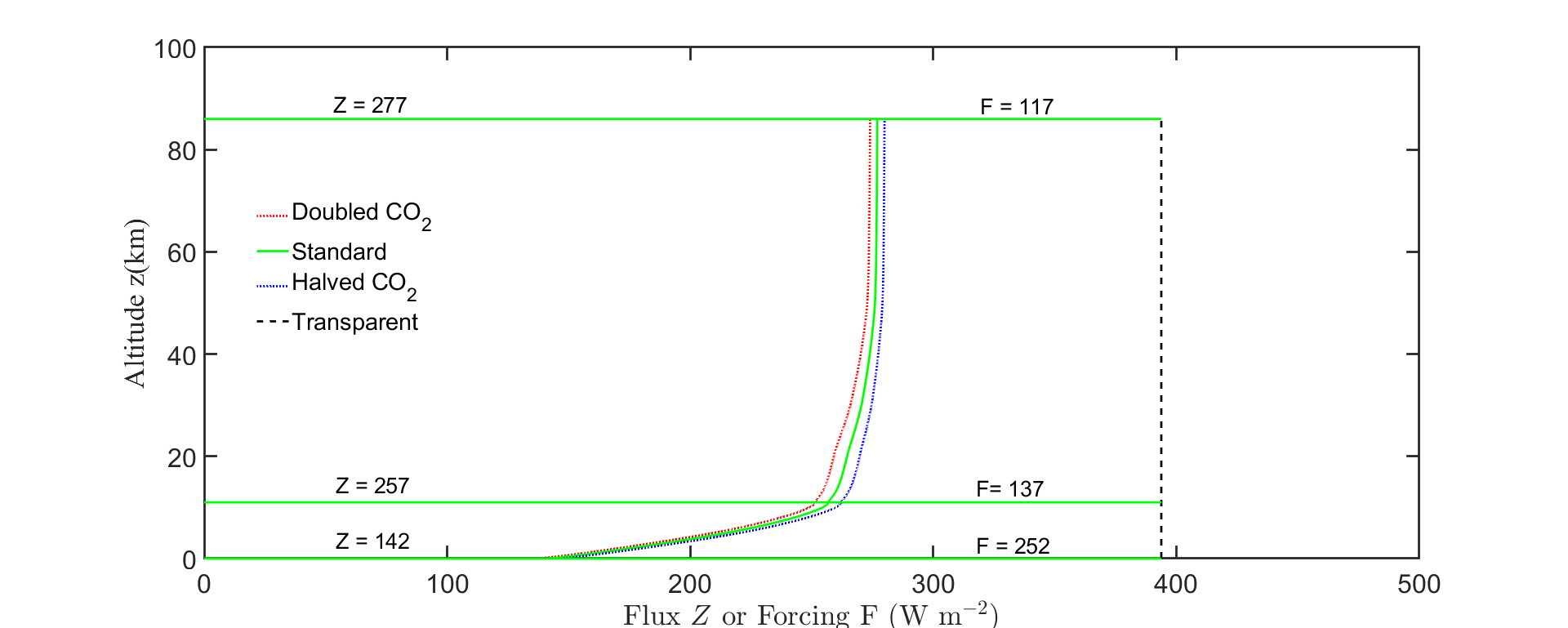} 
\caption{Altitude dependence of frequency integrated flux $Z$ of (\ref{b58}).  The flux for three concentrations of CO$_2$ are shown, the standard concentration, $C^{\{i\}}_{\rm sd}= 400$ ppm of Fig. \ref{GGNT}, twice and  half that value. The other greenhouse gases have the standard concentrations of Fig. \ref{GGNT}. The vertical dashed line is the flux $\sigma_{\rm SB}T_0^4=394$ W m$^{-2}$ for a transparent atmosphere with a surface temperature $T_0 = 288.7$ K. The forcings $F$ that follow from (\ref{b60}) at 0 km, 11 km and 86 km are 252, 137 and 117 W m$^{-2}$ respectively.
\label{ZzCM}}
\end{figure}

\begin{figure}[h]
\includegraphics[height=90mm,width=1.0\columnwidth]{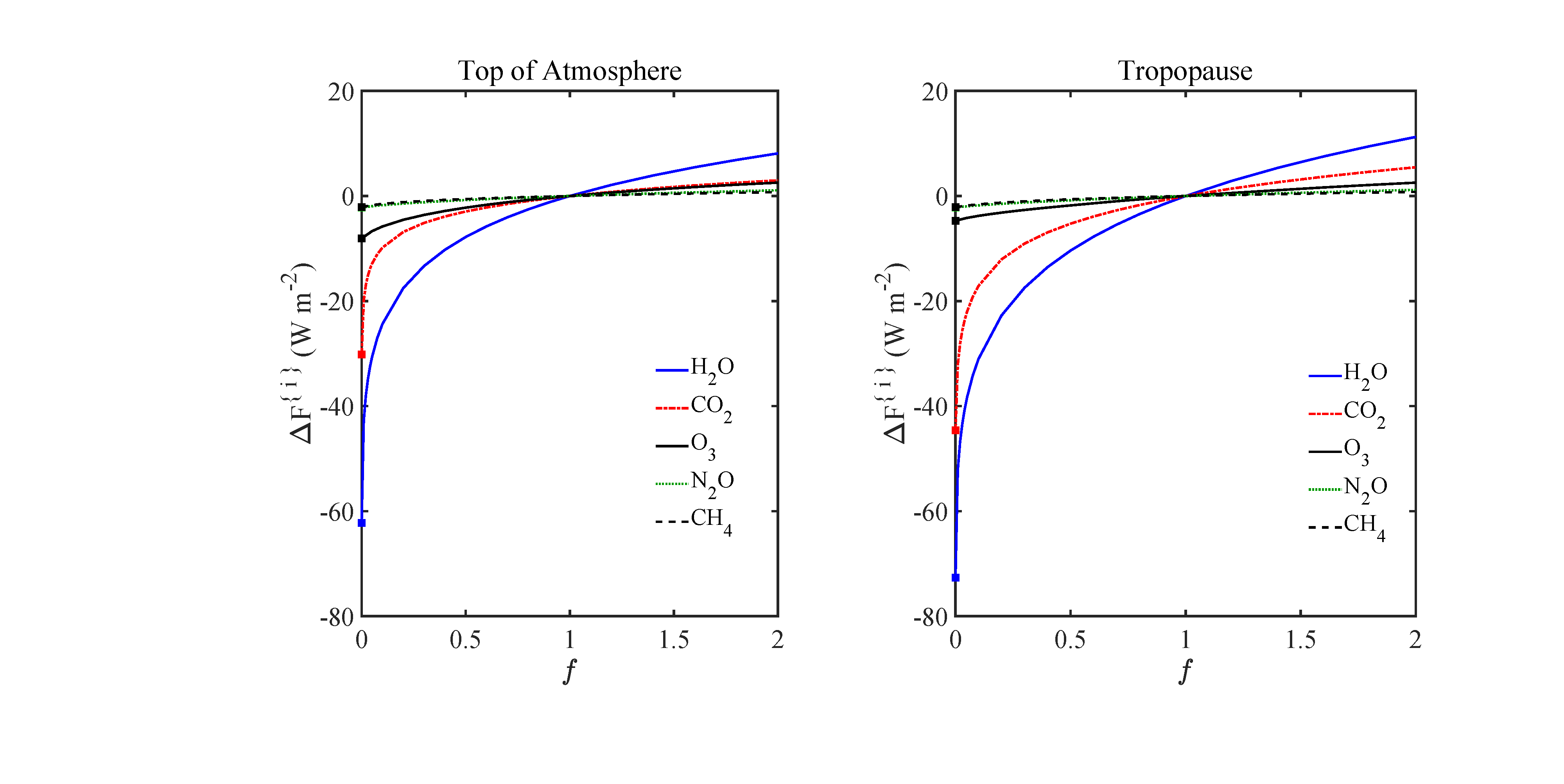} 
\caption{Dependence of partial forcing increments $\Delta F^{\{i\}}$ of (\ref{cdp18}) on greenhouse gas multiplicative factor, $f=N^{\{i\}}/N^{\{i\}}_{\rm sd}$ for H$_2$O, CO$_2$, O$_3$, N$_2$O and CH$_4$.  At the standard column densities, with $f=1$, the incremental forcings are well into the saturation regime, with $d\Delta F^{\{i\}}(1)/df< d\Delta F^{\{i\}}(0)/df $ for all 5 gases.  For the most abundant greenhouse gases, H$_2$O and CO$_2$, the saturation effects are extreme, with per-molecule forcing powers suppressed by four orders of magnitude at standard concentrations $(f=1)$ with respect to the low-concentration, optically thin limit $(f=0)$. For CO$_2$, the areas bounded by the green and black curves of Fig. \ref{CO2} give the values, $-\Delta F^{\{i\}}$ for $f=0$, while the areas bounded by the black and red curves give $\Delta F^{\{i\}}$ for $f=2$.  See the text and Table \ref{int} for more details.  The curves for CF$_4$ and SF$_6$ are shown in Fig. 7.  They are omitted here as they would appear as horizontal lines corresponding to nearly zero forcing. 
\label{DFDC}}
\end{figure}

Integrating spectral fluxes, $\tilde Z$, like those of Fig. \ref{CO2}, over all frequencies in accordance with (\ref{b58}) gives $Z$, the frequency integrated flux shown in Fig. \ref{ZzCM}.  A doubling of CO$_2$ concentration results in a 2.97 W/m$^2$ decrease in the top of the atmosphere flux.  

\begin{figure}[h]
\includegraphics[height=90mm,width=1.0\columnwidth]{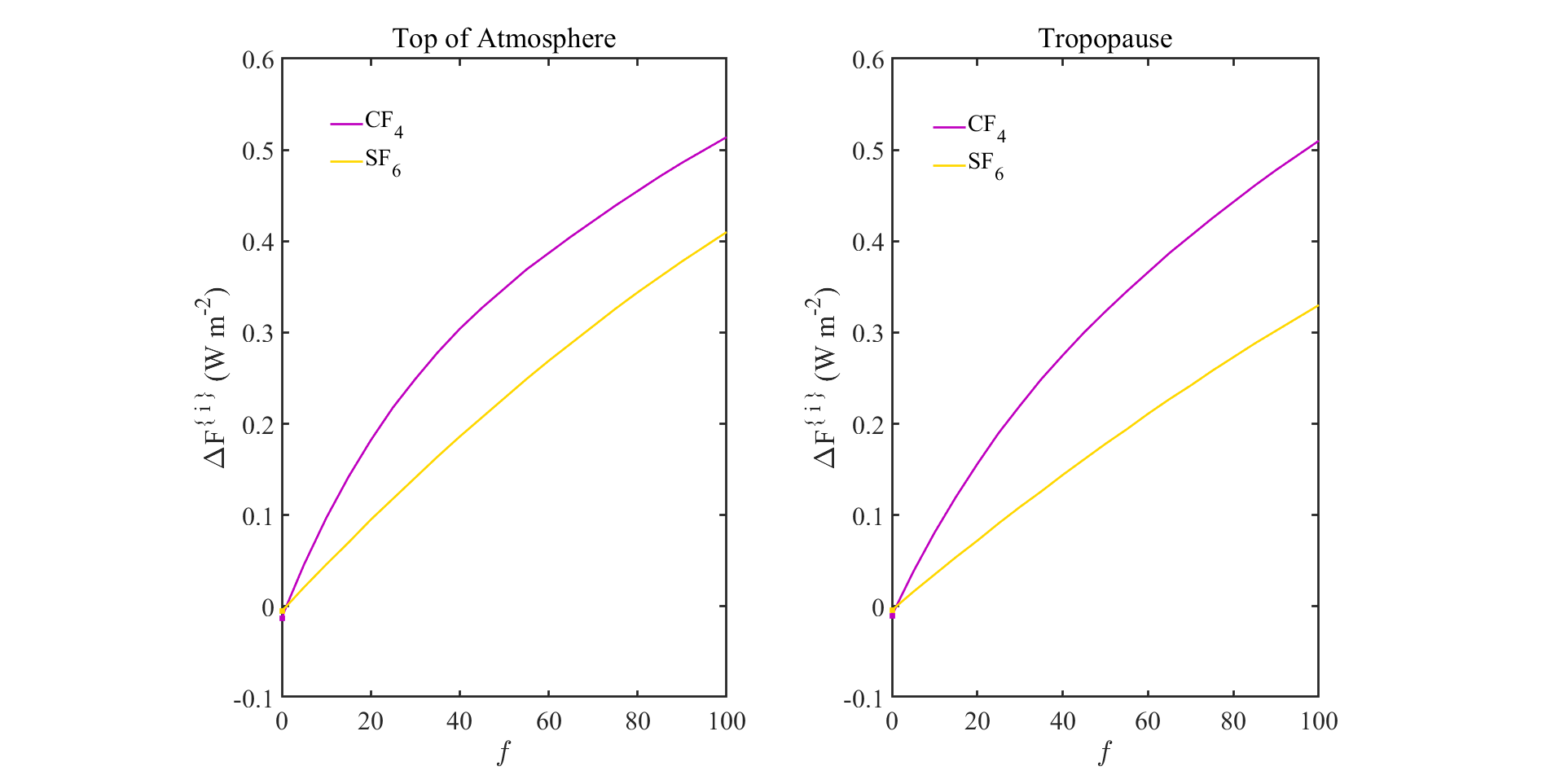} 
\caption{Dependence of partial forcing increments $\Delta F^{\{i\}}$ of (\ref{cdp18}) for CF$_4$ and SF$_6$ on greenhouse gas multiplicative factor, $f=N^{\{i\}}/N^{\{i\}}_{\rm sd}$.  This is similar to Fig. 6 but the range of $f$ values has been increased to 100. \label{DFDC2}}
\end{figure}

\section{Concentration Dependence of Forcing\label{cdp}}

\begin{table}
\begin{center}
\begin{tabular}{|c| r r|r r|r r|r r|}
\hline
Molecule &\multicolumn{2}{c|}{$F^{\{i\}}_{\rm sd}(z)$}  
&\multicolumn{2}{c|}{$\Delta F^{\{i\}}(z,0)$}
&\multicolumn{2}{c|}{$\Delta F^{\{i\}}(z,1/2)$}
&\multicolumn{2}{c|}{$\Delta F^{\{i\}}(z,2)$} \\

$i$& $z_{\rm tp}$ &$z_{\rm mp}$ &$z_{\rm tp}$ &$z_{\rm mp}$ 
&$z_{\rm tp}$ &$z_{\rm mp}$ &$z_{\rm tp}$ &$z_{\rm mp}$ \\ [0.5ex]
\hline\hline

H$_2$O&81.6&71.6&-72.6&-62.2&-10.4&-7.79&11.2&8.09\\
\hline
CO$_2$&52.4&38.9&-44.6&-30.2&-5.26&-2.96&5.48 &2.97\\
\hline
O$_3$&6.15&10.6&-4.70 &-8.09 &-1.80 &-2.22 &2.55 &2.55\\
\hline
N$_2$O&4.44&4.74&-2.16&-2.18&-0.81&-0.78&1.17 &1.09\\
\hline
CH$_4$&4.26&4.46&-2.12&-2.11&-0.63&-0.59&0.79 &0.74\\
\hline
CF$_4$ &0.034  &0.048  &-0.010 &-0.013 &-0.0050 &-0.0064 &0.0099  &0.012\\
\hline
SF$_6$ &0.0048  &0.0067 &-0.0041 &-0.0055 &-0.0020 &-0.0027 &0.0040  &0.0054 \\

\hline\hline
$\sum_i $&148.7&130.1&-126.2&-104.8&&&&\\
\hline\hline
$F_{\rm sd}(z)$&137&117&137&117&&&&\\
\hline
\end{tabular}
\end{center}

\caption{Partial forcings $F^{\{i\}}_{\rm sd}(z)$ of (\ref{cdp12}) and partial forcing increments $\Delta F^{\{i\}}(z,f)$
of (\ref{cdp18}), all in units of W m$^{-2}$, at the altitudes $z_{\rm tp}= 11$ km of the tropopause and $z_{\rm mp}= 86$ km of the mesopause. The last row contains the forcings $F_{\rm sd}(z)$ of (\ref{cdp10}), shown in Fig. \ref{ZzCM},
when all greenhouse molecules are present simultaneously at their standard column densities $\hat N^{\{i\}}_{\rm sd}$.
Because of the overlapping absorption bands, $\sum_i F^{\{i\}}_{\rm sd}(z)>F_{\rm sd}(z)$, and $-\sum_i \Delta F^{\{i\}}(z,0)<F_{\rm sd}(z)$. \label{int}}
\end{table}

The frequency integrated forcing, $F$, of (\ref{b60}) depends on the altitude $z$ and on the
column densities of the seven greenhouse gases given in Table \ref{acd}.
\begin{equation}
F=F(z,\hat N^{\{1\}},\ldots,\hat N^{\{7\}}).
\label{cdp8}
\end{equation}
We assume the temperature $T$ and densities $N^{\{i\}}$ have the same altitude profiles as shown in Fig. \ref{GGNT}.  An important special case of (\ref{cdp8}) is the forcing, $F_{\rm sd}$, when each greenhouse gas $i$ is present at its standard column density $\hat N^{\{i\}}_{\rm sd}$ of Table \ref{acd},
\begin{equation}
F_{\rm sd}(z)=F(z,\hat N^{\{1\}}_{\rm sd},\ldots,\hat N^{\{7\}}_{\rm sd}).
\label{cdp10}
\end{equation}
A second special case of (\ref{cdp8}) is the hypothetical, per molecule standard forcing, $F^{\{i\}}_{\rm sd}$, when the atmosphere contains only molecules of type $i$ at their standard column density, $\hat N^{\{i\}}=\hat N^{\{i\}}_{\rm sd}$, and the concentrations of the other greenhouse vanish, $\hat N^{\{j\}}=0$ if $j\ne i$,
\begin{equation}
F^{\{i\}}_{\rm sd}(z)=F(z,0,\ldots,0,\hat N^{\{i\}}_{\rm sd}, 0,\ldots,0).
\label{cdp12}
\end{equation}

We define the forcing power per added molecule as
\begin{equation}
P^{\{i\}}(z,\hat N^{\{1\}},\ldots,\hat N^{\{n\}})=\frac{\partial F}{\partial \hat N^{\{i\}}}.
\label{cdp14}
\end{equation}
The densities of greenhouse gases $j$ with $j\ne i$ are held constant in the
partial derivative of (\ref{cdp14}).  If the units of $F$ are taken to be W m$^{-2}$ and the units of $\hat N^{\{i\}}$ are taken to be molecules m$^{-2}$, then the units of $P^{\{i\}}$ will be W molecule$^{-1}$.

We define a finite forcing increment for the $i$th type of greenhouse molecule as
\begin{equation}
\Delta F^{\{i\}}(z,f)=F(z,\hat N^{\{1\}}_{\rm sd},\ldots,\hat N^{\{i-1\}}_{\rm sd},f\hat N^{\{i\}}_{\rm sd}, \hat N^{\{i+1\}}_{\rm sd},\ldots,
\hat N^{\{n\}}_{\rm sd})-F_{\rm sd}.
\label{cdp18}
\end{equation}
Differentiating (\ref{cdp18}) with respect to $f$ we find
\begin{equation}
\frac{\partial \Delta F^{\{i\}}}{\partial f}(z,f)=\hat N^{\{i\}}_{\rm sd}P^{\{i\}}_{\rm sd}(z,f),
\label{cdp20}
\end{equation}
where $P^{\{i\}}_{\rm sd}(z,f)$ is the forcing power per additional molecule of type $i$ when these molecules have the column density $\hat N^{\{i\}}=f\hat N^{\{i\}}_{\rm sd}$ and all other types of greenhouse molecules have their standard column densities.

The forcing increments (\ref{cdp18}) for the greenhouse gases considered in this paper are shown as a function of $f$ in Figs. \ref{DFDC} and \ref{DFDC2}. Forcing increments are also tabulated at representative altitudes $z$ and multiplicative factors $f$ in Table \ref{int}.  At both the top of the atmosphere and at the tropopause, we see that the forcing increment (\ref{cdp18}) is largest for abundant water molecules, H$_2$O, and is relatively small for the much more dilute greenhouse gases CH$_4$, N$_2$O, CF$_4$ and SF$_6$.  The incremental forcings are all in the saturation regime, with $\partial \Delta F^{\{i\}}/\partial f$ diminishing with increasing $f$.

In Table \ref{int}, the forcing decrements from removing H$_2$O, CO$_2$, O$_3$, N$_2$O and CH$_4$, $-62.2$, $-30.2$, $-8.1$, $-2.2$ and $-2.1$  W m$^{-2}$, are reasonably close to those calculated by Zhong and Haigh \cite{Zhong2013}. In their Table 1 they cite forcing decrements at the top of the atmosphere of $-70.6$, $-25.5$, $-7.0$, $-1.8$ and $-1.7$ W m$^{-2}$.  Zhong and Haigh seem to have taken the concentrations of N$_2$O and CH$_4$ to be independent of altitude. The altitude dependence of Fig. $\ref{GGNT}$ were used in our calculations.

\begin{table}
\begin{center}
\begin{tabular}{|c|c| c c|| c  c|c|}
\hline
& &\multicolumn{5}{c|} {$\Delta F^{\{i\}}(z,f)  \hbox{ in  W m}^{-2}$ }\\
\hline
Molecule& &\multicolumn{2}{c||}{This Work} &\multicolumn{3}{c|}{Previous Work}\\

$i$&$f$&$z_{\rm tp}$ &$z_{\rm mp}$ & $z_{\rm tp}$ & $z_{\rm mp}$ &Reference\\
\hline\hline
H$_2$O &1.06 &0.91 &0.67 &1.4 &1.1 &\cite{Collins2006} \\ \hline
CO$_2$ &2    &5.48 &2.97 &5.5 &2.8 &\cite{Collins2006} \\ \hline
O$_3$  &1.1  &0.30 &0.33 &    &    & \\ \hline
N$_2$O &2    &1.17 &1.09 &1.3   &1.2 &\cite{Collins2006} \\ \hline
CH$_4$ &2    &0.79 &0.74 &0.6 &0.6 &\cite{Collins2006} \\ \hline
CF$_4$ &2    &0.0099 &0.012 &0.0088    &    &\cite{Hurley} \\ 
       &100  &0.51&0.51&    &    &    \\ \hline
SF$_6$ &2    &0.0040 &0.0054 &0.0056 & &\cite{Hodnebrog}\\
       &100  &0.33&0.41&    &    &      \\
\hline
\end{tabular}
\end{center}
\caption{Comparison of the forcing increments $\Delta F^{\{i\}}(z,f)$ in column 5 of Table 2 and previous work at the altitude $z_{\rm tp} = 11$ km of the tropopause and $z_{\rm mp} = 86$ km of the mesopause.  For H$_2$O, the relative increase, $f=1.06$, of the column density is that predicted by the Clausius Clapeyron equation for a 1 K temperature increase. For CF$_4$ and SF$_6$, saturation effects are evident at $f=100$ as shown in Fig. 7.  Note that if there were no saturation $\Delta F^{\{i\}}(z,100) = 100 \Delta F^{\{i\}}(z,2)$ since $\Delta F^{\{i\}}(z,1)=0$.
\label{int2}}
\end{table}
The forcing increments for the 5 naturally occurring greenhouse gases given  in Table \ref{int2} are comparable to those calculated by others. We give the increments  $\Delta F^{\{i\}}(z,f)$ calculated by Collins et al \cite{Collins2006}, as
estimated from their Tables 2 and 8.  These are the results of averaging five separate line by line calculations.
Our forcing increment for doubling CF$_4$ is reasonably close to that obtained using the radiative efficiency of 0.102 W m$^{-2}$ ppb$^{-1}$ found by Hurley et al \cite{Hurley}.  They predict the radiative forcing will decrease by up to 40\% at higher CF$_4$ concentrations as a result of overlapping absorption due to CH$_4$, H$_2$O and N$_2$O which is consistent with our result for $f=100$.  

The forcing increment for doubling SF$_6$ was found using both the VAMDC and HITRAN datsets.  The VAMDC dataset lists about 10\% more lines than HITRAN and the values of the individual line strengths are about 6\% larger.  
The resulting forcing increments shown in Tables 2 and 3 are about 20\% larger than obtained using the HITRAN data. 
Our tropopause forcing increment for doubling SF$_6$, 0.40 W/m$^{-2}$ ppb$^{-1}$, is significantly lower than that given in the review by Hodnebrog et al \cite{Hodnebrog}.  Their result is an average of several studies that used absorption cross sections giving results ranging from 0.49 to 0.68 W m$^{-2}$ ppb$^{-1}$.  A recent result by Kovacs et al \cite{Kovacs}, calculated radiative efficiencies of 0.77 and 0.50 W m$^{-2}$ ppb$^{-1}$ for clear and all sky cases, respectively.  Our work and that of Kovacs both used the same SF$_6$ altitudinal profile shown in Fig. 1.  Kovacs also found the integrated cross sections for infrared absorption of $2.02 \times 10^{-16}$ cm$^2$ for the 925-955 cm$^{-1}$ band.  This is close to that found recently by Harrison who examined CF$_4$ and SF$_6$ absorption cross sections \cite {HarrisonCF4, HarrisonSF6}.  Their result for the frequency-integrated cross section was assumed to be independent of temperature in contrast to our results as discussed in Section 6.

The three mesopause flux increments $\Delta F^{\{i\}}$ in the fourth column of Table \ref{int2} for doubled concentrations of CO$_2$, N$_2$O and CH$_4$ sum to 4.8 W m$^{-2}$.  The calculated flux increment from simultaneously doubling CO$_2$, N$_2$O and CH$_4$ is the slightly smaller value, $\Delta F = 4.7$ W m$^{-2}$.
Similarly, the four mesopause flux increments $\Delta F^{\{i\}}$ in the fourth column of Table \ref{int2} for doubled concentrations of CO$_2$, N$_2$O and CH$_4$ as well as a factor of $f=1.06$ increase of H$_2$O concentration sum to 5.5 W m$^{-2}$.
The calculated flux increment from simultaneously doubling CO$_2$, N$_2$O and CH$_4$, and increasing the H$_2$O concentrations by a factor of $f=1.06$, is the slightly smaller value 5.3 W m$^{-2}$.  The ``whole" is  less than the sum of the parts, because of the interference of greenhouse gases that absorb the same infrared frequencies.

\begin{figure}[h]
\includegraphics[height=90mm,width=1.0\columnwidth]{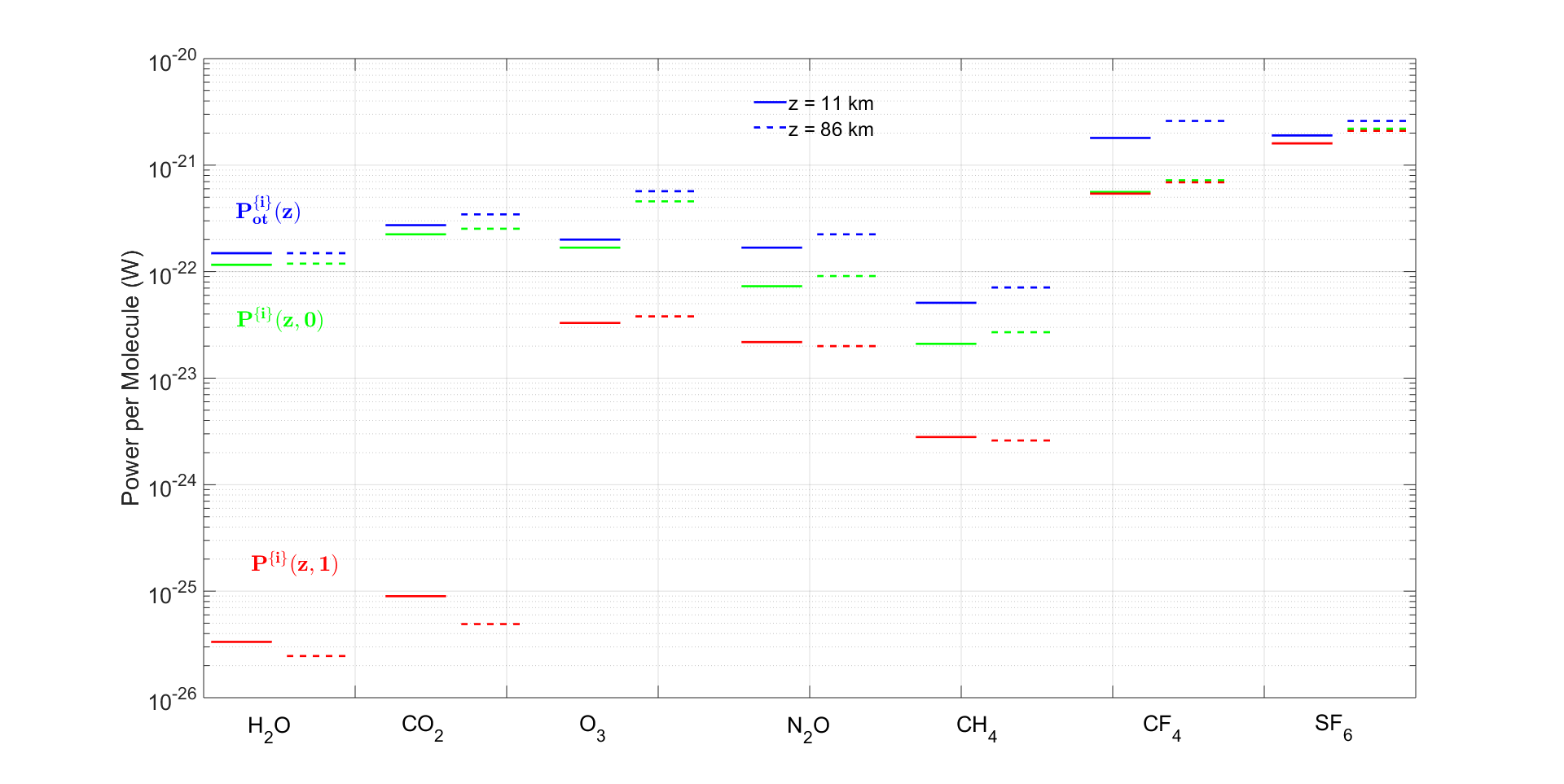} 
\caption{A graphical display of the per-molecule forcing powers of Table \ref{dPr}. At standard column densities the (red) powers, $P^{\{i\}}_{\rm sd}(z,1)$, for H$_2$O and CO$_2$ are suppressed by four orders of magnitude from their values in the optically thin limit (blue) where the powers are $P^{\{i\}}_{\rm ot}(z)$. This is due to strong saturation of the absorption bands. Saturation effects (difference between the blue and red lines) are much less for the minor gases, O$_3$, N$_2$O and CH$_4$. The green lines are the powers per molecule, $P^{\{i\}}_{\rm sd}(z,0)$, of the $i$th greenhouse gas in its low-concentration limit, but when the forcing power is suppressed by other gases at their standard densities.  Interference effects (difference between the blue and green lines) are more pronounced for N$_2$O and CH$_4$ than for H$_2$O and CO$_2$. Fig. \ref{Fig2} shows the strongest bands of O$_3$ overlap little with those of other greenhouse molecules, minimizing interference effects.  Note that for CF$_4$ and SF$_6$, the red and green curves nearly overlap.
\label{Pbar}}
\end{figure}

Table \ref{dPr} summarizes the forcing powers (\ref{cdp14}) per additional molecule in units of $10^{-22}$ W at
the tropopause altitude, $z_{\rm tp}=11$ km and at the  mesopause altitude, $z_{\rm mp}=86$ km. The surface temperature was $T_0=288.7$ K,
and the altitude profiles of temperature and number density were those of Fig. $\ref{GGNT}$.  The second column lists the forcing powers in the optically thin limit, $P^{\{i\}}_{\rm ot}(z)$, of (\ref{ot4}).  These numbers agreed with the results of dividing the forcing in the optical thin limit by the column density.  The numbers of the third column are forcing powers $P^{\{i\}}_{\rm sd}(z,0)$ from (\ref{cdp20}) for an atmosphere that previously had no molecules of type $i$ (so $\hat N^{\{i\}}=0$) but all other greenhouse molecules had standard concentrations, $\hat N^{\{j\}}=\hat N^{\{j\}}_{\rm sd}$ if $j\ne i$. The forcings of the third column are less than those of the second because of interference between absorption by different greenhouse gases. The numbers in the fourth column are the forcing powers $P^{\{i\}}_{\rm sd}(z,1)$ from (\ref{cdp20}) when a single molecule of type $i$ is added to an atmosphere that previously had standard  densities for all greenhouse gases, $\hat N^{\{j\}}=\hat N^{\{j\}}_{\rm sd}$. Saturation of the absorption suppresses the per-molecule forcing by about four orders of magnitude for the abundant greenhouse gases H$_2$O and CO$_2$. Saturation causes less drastic suppression of per-molecule forcings for the less abundant O$_3$, N$_2$O and CH$_4$.  Saturation is much less for CF$_4$ and SF$_6$ because the concentrations of these gases are several orders of magnitude smaller than any of the naturally occurring greenhouse gases.  The forcing powers per molecule are summarized graphically in Fig. \ref{Pbar}.

\begin{table}
	\begin{center}
		\begin{tabular}{|c| c c | c c| c c |}
			\hline
			Molecule&\multicolumn{2}{c|}{$P^{\{i\}}_{\rm ot}(z)$}&\multicolumn{2}{c|}{$P^{\{i\}}_{\rm sd}(z,0)$}&\multicolumn{2}{c|}{$P^{\{i\}}_{\rm sd}(z,1)$}\\
		
$i$& $z_{\rm tp}$ & $z_{\rm mp}$  & $z_{\rm tp}$ & $z_{\rm mp}$   & $z_{\rm tp}$  & $z_{\rm mp}$  \\ [0.5ex]
			\hline\hline
			H$_2$O&1.50&1.50&1.2&1.2& $3.3\times 10^{-4}$ & $2.5\times 10^{-4}$ \\
			\hline
			CO$_2$&2.76&3.49&2.2&2.5&$9.0\times 10^{-4}$&$4.9\times 10^{-4}$\\
			\hline
			O$_3$&2.00&5.71&1.7&4.6&$3.3\times 10^{-1}$&$3.8\times 10^{-1}$\\
			\hline
			N$_2$O&1.70&2.27&0.73&0.91&$2.1\times 10^{-1}$& $2.0\times 10^{-1}$\\
			\hline
			CH$_4$&0.52&0.72&0.21&0.27&$2.8\times 10^{-2}$&$2.6\times 10^{-2}$\\
			\hline
			CF$_4$&19.5 &28.2 &5.6 &7.2 &5.4 &6.9\\
			\hline
			SF$_6$&23.3 &32.1 &19 &26 &19 &26\\
			\hline
		\end{tabular}
	\end{center}
	\caption{Forcing powers (\ref{cdp14}) per additional molecule in units of $10^{-22}$ W at the altitude $z_{\rm tp} = 11$ km of the tropopause and $z_{\rm mp} = 86$ km of the mesopause. The surface temperature was $T_0=288.7$ K and the altitude profiles of temperature and number density were those of Fig. $\ref{GGNT}$.  $P^{\{i\}}_{\rm ot}(z)$ of (\ref{ot6}) is for the optically-thin limit.  $P^{\{i\}}_{\rm sd}(z,0)$ from (\ref{cdp20}) is for an atmosphere that previously had no molecules of type $i$ (so $\hat N^{\{i\}}=0$) but all other greenhouse molecules had standard concentrations.
		$P^{\{i\}}_{\rm sd}(z,1)$ from (\ref{cdp20}) is for a single molecule of type $i$ added to an atmosphere that previously had standard densities for all greenhouse gases.\label{dPr}}
\end{table}

We now consider the optically thin limit, where the concentrations of greenhouse gases are sufficiently low that the optical depths $\tau$ of (\ref{b10}) will be small, $\tau\ll 1$, for all frequencies $\nu$ and at all altitudes $z$.  The frequency integral of the spectral forcing (\ref{b58}) at altitude $z$ can then be written as

\begin{eqnarray}
F_{\rm ot}(z)&=&\sum_i\hat N^{\{i\}} P^{\{i\}}_{\rm ot}(z)
\label{ot2}
\end{eqnarray}

\noindent where the forcing power per molecule of type $i$ is

\begin{equation}
P^{\{i\}}_{\rm ot}(z)=\frac{1}{2}\int_0^z dz'\frac{N^{\{i\}'}}{\hat N^{\{i\}}}\left[\Pi^{\{i\}}(T',T_0)-\Pi^{\{i\}}(T',T')\right]
+\frac{1}{2}\int_z^{\infty} dz'\frac{N^{\{i\}'}}{\hat N^{\{i\}}}\Pi^{\{i\}}(T',T').
\label{ot4}
\end{equation}

\noindent Here $N^{\{i\}'} = N^{\{i\}}(z')$,  $T'=T(z')$ and $T_0=T(0)$. 
The mean power absorbed by a greenhouse gas molecule of temperature $T$  from thermal equilibrium radiation of temperature $T'$ is

\begin{equation}
\Pi^{\{i\}}(T,T')= 4\pi\sum_{ul} S_{ul}^{\{i\}}(T)  \tilde B(\nu_{ul},T').
\label{ot6}
\end{equation}

\noindent For $z'>z$ we see from (\ref{ot4}) that the $dz'\, N'$ molecules in the altitude interval $z'$ to $z'+dz'$ each emit the power $\Pi(T',T')$, of which half goes to outer space and half goes down through the reference plane at altitude $z$, diminishing the net flux through the reference plane by $\Pi(T',T')/2$.  Molecules above the reference plane can only cause positive forcing unlike molecules below the reference plane which can cause either positive or negative forcing.

For the special case of $T'=T$ we can substitute (\ref{lbl24}) into (\ref{ot6}) to find
\begin{eqnarray}
\Pi^{\{i\}}(T,T)&=&\sum_{ul}W_u^{\{i\}}(T)\Gamma_{ul}^{\{i\}}E_{ul}^{\{i\}}.
\label{ot8}
\end{eqnarray}
Since we are considering a single isotopologue, we have set $\eta_u=1$ in (\ref{lbl24}).  The three factors in the summed terms of (\ref{ot8}) are  the probability $W_u^{\{i\}}(T)$ to find the molecule in the upper state $u$, the radiative decay rate $\Gamma_{ul}^{\{i\}}$ from the upper level $u$ to the lower level $l$ and the energy $E_{ul}^{\{i\}}$ of the emitted photon.  This is obviously the total power radiated by a molecule of temperature $T$.  For a molecule of temperature $T$ in thermal equilibrium with radiation of the same temperature, the radiative power absorbed by the molecule is equal to the spontaneous radiative power it emits. 

An important check of our work was to use (\ref{ot4}) to calculate the forcing power per molecule for the optically thin limit.  The results displayed in column 2 of Table 4 agreed with that found by dividing the radiative forcing in the limit of zero gas concentration, by the column density.

\section{Harmonic Oscillators}
The emission and absorption of thermal radiation by greenhouse-gas molecules usually decreases or increases the quantized vibrational energy, with the notable exception of the low-frequency transitions of H$_2$O, for which rotational quantum numbers change, but there is no change of the vibrational quantum numbers.  Linear molecules with $N$ atoms, like CO$_2$ or N$_2$O, have  $3N-5$ vibrational modes whereas nonlinear molecules, like H$_2$O or CH$_4$,  have $3N-6$ vibrational modes \cite{Landau}.  For molecules of sufficient symmetry, several independent modes may have the same frequency $\nu_i$. They correspond to vibrations along $d_i$ mutually perpendicular unit vectors ${\bf x}_j$.
The possible degeneracies $d_i$ of the $i$th set of equal-frequency modes are
\begin{equation}	
d_i = 1, 2, 3. \label{so2}
\end{equation}
Summing over the degeneracies, we must have 
\begin{eqnarray}	
\sum_i d_i &=& 3N-5,\quad\hbox{for linear molecules}.\label{so4}\\
\sum_i d_i &=& 3N-6,\quad\hbox{for nonlinear molecules}.\label{so6}\\
\end{eqnarray}

The bending modes of the linear molecules  CO$_2$ and N$_2$O, with resonance frequencies  $\nu_2$ of 667 and 588  cm$^{-1}$ respectively, can be represented as two-dimensional oscillators with degeneracy $2$.  The bending modes of CH$_4$ and CF$_4$, with resonance frequencies 1311 and 631 cm$^{-1}$ respectively, are well approximated as three-dimensional oscillators, with degeneracy $3$.  The asymmetric stretch modes of CO$_2$ and N$_2$O with resonance frequencies 2349 and 2224 cm$^{-1}$ respectively, can be represented as one-dimensional oscillators, with single degeneracy. 
 
Molecular rotation makes little difference to the the total radiated  power, but it  spreads the emitted frequencies over a band centered on the vibration frequency $\nu_i$ of the non-rotating molecule. Examples are shown in Fig. \ref{Fig2}. The P and R branches of the bands are created by the rotation of the radiation pattern about an axis that is not parallel to the vibration axis.  Q branches have a relatively narrow spectrum because the rotation and vibration axes of the molecule are nearly parallel. The permanently bent, asymmetric-top molecules, H$_2$O, and O$_3$, have especially complicated rotational sidebands.
\subsection{Quantum Mechanics}
Consider the effective mass of the $i$th vibrational mode of a molecule to be $m_i$, and let the restoring-force constant be $k_i$.  We can write the Hamiltonian for the vibrational energy of the $i$th mode of the non-rotating molecule as
\begin{equation}	
H_i=\sum_{j=1}^{d_i}H_{ij}.\label{so8}
\end{equation}
The Hamiltonian describing vibrations of the $i$th mode along the $j$th spatial axis is
\begin{eqnarray}	
H_{ij}&=&-\frac{\hbar^2}{2m_i}\frac{\partial^2}{\partial x_{ij}^2}+\frac{k_i}{2}x_{ij}^2-\frac{\hbar\omega_i}{2}\nonumber\\
&=&\frac{\hbar\omega_i}{2}\left(-\frac{\partial^2}{\partial \xi_{ij}^2}+\xi_{ij}^2-1\right)\nonumber\\
&=&\frac{\hbar\omega_i}{2}\left(a_{ij} a_{ij}^{\dag}+a_{ij}^{\dag}a_{ij}-1\right)\nonumber\\
&=&\hbar\omega_i\,a_{ij}^{\dag}a_{ij}.
\label{so10}
\end{eqnarray}
Here $x_{ij}$ is the displacement coordinate along the $j$th axis. For convenience,  
  we have subtracted the zero-point vibrational energy, $\hbar \omega_i/2$ from the Hamiltonian. The resonant angular frequency, in radians per second, is
\begin{equation}	
\omega_i=\sqrt{\frac{k_i}{m_i}}=2\pi c\,\nu_i.
\label{so12}
\end{equation}
The dimensionless displacements are
\begin{equation}	
\xi_{ij}=\frac{x_{ij}}{X_i},
\label{so14}
\end{equation}
where the characteristic length  is
\begin{equation}	
X_i=\left(\frac{\hbar^2}{m_i k_i}\right)^{1/4}.
\label{so16}
\end{equation}
The annihilation and creation operators for excitations along the $j$th spatial axis are
\begin{equation}	
a_{ij}=\frac{1}{\sqrt{2}}\left(\xi_{ij}+\frac{\partial}{\partial \xi_{ij}}\right) \quad\hbox{and}\quad
a_{ij}^{\dag}=\frac{1}{\sqrt{2}}\left(\xi_{ij}-\frac{\partial}{\partial \xi_{ij}}\right).
\label{so18}
\end{equation}
The non-zero commutators, which have the values
\begin{equation}	
a_{ij}a_{ij}^{\dag} -a_{ij}^{\dag}a_{ij}=1,
\label{so20}
\end{equation}
can be used to derive the last line of (\ref{so10}).

The energy basis states, $\phi_{{\bf n}_{i}}$, of the molecule are the eigenfuctions of the Hamiltonian (\ref{so8})
\begin{equation}	
H_{i}\phi_{{\bf n}_{i}} =E_{n_i}\phi_{{\bf n}_{i}}.
\label{so22}
\end{equation}
We have labeled the energy eigenstates with the spatial vector
\begin{equation}	
{\bf n}_{i}=n_{i1}{\bf x}_1+n_{i2}{\bf x}_2+\cdots+n_{id_i}{\bf x}_{d_i}
\label{so23}
\end{equation}
The total number $n_i$ of excitation quanta of the state $\phi_{{\bf n}_{i}}$ is
\begin{equation}	
n_i= \sum_{j=1}^{d_i} n_{ij}=\sum_{j=1}^{d_i} {\bf n}_{i}\cdot{\bf x}_j
={\bf n}_i\cdot{\bf d}_i \, ,
\label{so26}
\end{equation}
where  the vector ${\bf d}_i$ is
\begin{equation}	
{\bf d}_{i}={\bf x}_1+{\bf x}_2+\cdots+{\bf x}_{d_i}.
\label{so27}
\end{equation}
The energy eigenvalue of (\ref{so22}) is
\begin{equation}	
E_{n_i}=\hbar\omega_i n_i.
\label{so28}
\end{equation}
A few examples may help to clarify the preceding discussion.
If the mode degeneracy is $d_i=3$, and there are $n_i=0$ vibrational excitation quanta, the possible values of $n_{ij}$ are
\begin{equation}
[n_{i1}, n_{i2},n_{i3}]= [0,0,0].
\label{so30}
\end{equation}
The  energy degeneracy is $g_i(0)=1$.
The ground state of a three-dimensional oscllator is a non-degenerate S state.

For $n_i=1$ the possible values of $n_{ij}$ are
\begin{equation}
[n_{i1}, n_{i2},n_{i3}]= [1,0,0],\quad[0,1,0],\quad[0,0,1].
\label{so32}
\end{equation}
The energy degeneracy is $g_i(1)=3$.
The independent states are linear combinations of the 3 azimuthal sublevels of a P state.

For $n_i =2$ the possible values of $n_{ij}$ are
\begin{equation}
[n_{i1}, n_{i2},n_{i3}]= [2,0,0],\quad[0,2,0],\quad[0,0,2],\quad[0,1,1],\quad[1,0,1],\quad[1,1,0].
\label{so34}
\end{equation}
The energy degeneracy  is $g_i(2)=6$. One of the states is a non-degenerate S state while the 5 other degenerate states are linear combinations of the 5 azimuthal sublevels of a D state.

Following  this line of reasoning, we see that for a mode degeneracy $d_i$, and for $n_i$ vibrational excitation quanta, the energy degeneracy $g_i(n_i)$ is given by the formula

\begin{equation}
g_i(n_i)=\frac{(n_i+d_i-1)!}{n_i!(d_i-1)!}=\sum_{{\bf n}_i}1,\quad\hbox{where} \quad{\bf n}_i\cdot{\bf d}_i =n_i .
\label{so36}
\end{equation}
The energy basis states of (\ref{so22}) can be written as
\begin{equation}	
\phi_{{\bf n}_{i}}=\frac{(a_1^{\dag})^{n_{i1}}(a_2^{\dag})^{n_{i2}}\cdots(a_d^{\dag})^{n_{id_i}}}{\sqrt{n_{i1}! n_{i2}!\cdots n_{id_i}!}}\phi_{{\bf 0}}.
\label{so38}
\end{equation}
The non-degenerate, ground-state wave function is
\begin{equation}	
\phi_{\bf 0}=\frac{e^{-(\xi_{i1}^2+\xi_{i2}^2+\cdots+\xi_{id_i}^2)/2}}{\pi^{d_i/4}}.
\label{so40}
\end{equation}
The commutation relation (\ref{so20}) implies that
\begin{equation}	
a_{ij}\phi_{{\bf n}_i}=\sqrt{n_{ij}}\,\phi_{{\bf n}_{i}-{\bf x}_j},
\label{so42}
\end{equation}
and also
\begin{equation}	
a_{ij}^{\dag}\phi_{{\bf n}_i}=\sqrt{n_{ij}+1}\,\phi_{{\bf n}_{i}+{\bf x}_j}.
\label{so44}
\end{equation}
The numbers of quanta along the vibrational axes of the ``raised'' and ``lowered'' states are defined by
\begin{equation}	
{\bf n}_{i}\pm {\bf x}_j=n_{i1}{\bf x}_1+\cdots+( n_{ij}\pm 1){\bf x}_j+\cdots+ n_{id_i}{\bf x}_{d_i}.
\label{so46}
\end{equation}
\subsection{Radiation}
The electric-dipole-moment operator of the molecule is defined by
\begin{eqnarray}	
{\bf M}_i&=&M_i\sum_{j=1}^{d_i}\xi_{ij}{\bf x}_j\nonumber\\
&=&\frac{M_i}{\sqrt{2}}\sum_{j=1}^{d_i}(a_{ij}+a_{ij}^{\dag}){\bf x}_j.
\label{so48}
\end{eqnarray}
The moment amplitude $M_i$ is the product of the characteristic length of (\ref{so16}) and a characteristic charge $q_i$,
\begin{equation}	
M_i=q_iX_i.
\label{so50}
\end{equation}
We denote the matrix elements by
\begin{equation}
\langle\,{\bf n}_i\ |{\bf M}_i|\,{\bf n}_i'\ \rangle=\int_{-\infty}^{\infty}d\xi_{i1}\cdots\int_{-\infty}^{\infty}d\xi_{id_i}\ \phi^*_{{\bf n}_{i}}(\xi_{i1},\ldots,\xi_{id_i}){\bf M}_i\ \phi_{{\bf n}_{i}'}(\xi_{i1},\ldots,\xi_{id_i}).
\label{so52}
\end{equation}

When the molecule is approximated as a non-rotating, $d_i$-dimensional harmonic oscillator, $\nu_{ul}=\nu_i=\omega_i/2\pi c$, where the ``principal quantum number'' of the upper level $u$ is $n_i ={\bf n}_i\cdot{\bf d}_i$ and the principal quantum number of the lower level $l$ is $n_i'={\bf n}_i'\cdot {\bf d}_i=n_i-1$. The energies of the upper and lower levels are $E_u =\hbar\omega_i n_i$ and $E_l =\hbar\omega_i n_i'=E_u-\hbar\omega_i$. Then we can use (\ref{so48}), (\ref{so42}), (\ref{so44}) (\ref{so26}) and (\ref{so36}) to write the oscillator strength (\ref{lbl30}) as

\begin{eqnarray}
f_{ul}&=&\frac{4\pi\nu_i}{3c\,r_e\hbar\, g_i(n_i')}\sum_{{\bf n}_{i},{\bf n}_{i}'}\langle\,{\bf n}_{i}]\,{\bf M}_i|\,{\bf n}_{i}'\,\rangle\cdot
\langle \,{\bf n}_i'\,|{\bf M}_i|\,{\bf n}_i\,\rangle\nonumber\\
&=&\frac{2\pi\nu_i M_i^2}{3c\,r_e\hbar g_i(n_i')}\sum_{{\bf n}_i}\langle \,{\bf n}_{i}\,|\sum_{j=1}^{d_i} n_{ij}|\,{\bf n}_{i}\,\rangle\nonumber\\
&=&\frac{2\pi\nu_i M_i^2}{3c\,r_e\hbar \,g_i(n_i')}\sum_{{\bf n}_{i}} n_{i} \nonumber\\
&=&\frac{2\pi\nu_i M_i^2n_i g_i(n_i) }{3c\,r_e\hbar\, g_i(n_i')}.
\label{soo54}
\end{eqnarray}
Substituting (\ref{soo54}) into (\ref{lbl31}) we find the spontaneous decay rate
\begin{equation}
\Gamma_{n_i}=\Gamma_{ul}
=\frac{2\omega_i^3M_i^2n_i}{3c^3\hbar}.\label{soo56}
\end{equation}
 Multiplying (\ref{soo56}) by the photon energy, $\hbar\omega_i$, we find the power radiated by molecules at the frequency $\omega_i$ of the $i$th mode, with $n_i$ vibrational excitation quanta, is
\begin{equation}
P_i(n_i) = \hbar \omega_i \Gamma_{n_i}
=\frac{2\omega_i^4M_i^2n_i}{3c^3}.\label{soo58}
\end{equation}
The power (\ref{soo58}) is proportional to the number of vibrational excitation quanta $n_i$, to the fourth power of the vibration angular frequency $\omega_i$, and to the square of the transition moment $M_i$.  As expected from the {\it correspondence principle}, the quantum mechanical expression (\ref{soo58}) for the radiated power is closely analogous to  Larmor's classical formula for the power radiated by an oscillating dipole moment, $M=M_o\cos\omega t$ with $\ddot M=-\omega^2 M$,
\begin{equation}
P=\frac{2\ddot M^2}{3c^3}=\frac{2\omega^4 M^2}{3c^3}.\label{so59}
\end{equation}
\subsection{Statistical Mechanics}
In accordance with (\ref{lbl26}), the Boltzmann probability that a molecule at a temperature $T$ will have $n_i$ vibrational excitation quanta for the $i$th mode is
\begin{equation}
W_i(n_i) =\frac{g_i(n_i) e^{-\hbar \omega_i n_i /k_{\rm B}T}}{Q}.
\label{soo60}
\end{equation}
Here the vibrational partition function for the non-rotating molecule with $m$ different vibrational modes is
\begin{equation}
Q= Q_1 Q_2\cdots Q_m.
\label{so62}
\end{equation}
The partition function of the $i$th mode is
\begin{equation}
Q_i =\sum_{n_i=0}^{\infty}g_i(n_i) e^{-\hbar \omega_i n_i /k_{\rm B}T}.
\label{so64}
\end{equation}
From (\ref{soo58}) and (\ref{soo60}) we see that at a temperature $T$, the average power emitted by a molecule due to vibrations of the $i$th normal mode is
\begin{equation}
\langle P_i \rangle  
=\frac{2\omega_i^4M_i^2\langle n_i\rangle}{3c^3}.\label{so66}
\end{equation}
where the mean number of vibrational excitation quanta of the $i$th mode is
\begin{eqnarray}
\langle n_i \rangle&=&\sum_{n_i=0}^{\infty} n_i W_i(n_i)\nonumber\\
&=&\sum_{n_i=0}^{\infty}  \frac{n_i g_i(n_i) e^{-\hbar \omega_i n_i /k_{\rm B}T}}{Q}\nonumber\\
&=&\frac{k_{\rm B}T^2Q_i}{\hbar\omega_i Q}\frac{d}{dT}\ln Q_i
\label{so68}
\end{eqnarray}
%


The evaluation of the partition functions can be facilitated by noting the identity
\begin{equation}	
\frac{1}{(1-x)^{d_i}}
=\sum_{n_i=0}^{\infty} g_i(n_i) x^{n_i}.
\label{so70}
\end{equation}
where the energy degeneracy $g_i(n_i)$ of the $i$th mode, with $n_i$ excitation quanta, was given by (\ref{so36}). We can use (\ref{so70}) to write the vibrational partition function (\ref{so64}) of the $i$th mode as
\begin{equation}
Q_i=\sum_{n_i=1}^{\infty} g_i(n_i) e^{-\hbar \omega_i n_i/k_{\rm B}T}
=\frac{1}{(1-e^{-\hbar \omega_i/k_{\rm B}T})^{d_i}}
\label{so72}
\end{equation}
%
%
%
Substituting (\ref{so72}) into (\ref{so68}) we find
\begin{eqnarray}
\langle n_i \rangle&=&
\frac{ d_i}{(e^{\hbar \omega_i/k_{\rm B}T}-1)}\left(\frac{Q_i}{Q}\right)
\label{so76}
\end{eqnarray}
For many of the modes listed in Table 5 such as the CO$_2$ bending mode with vibrational frequency $\nu_i = 667 $ cm$^{-1}$, the factor $Q_i/Q$ hardly differs from 1 over the range of temperatures observed in Earth's atmosphere.  But for other modes such as the CO$_2$ asymmetric stretch mode, with the vibrational frequency $\nu_i  = 2349$ cm$^{-1}$, the factor $Q_i/Q$ is several percent less than 1.  For the infrared active modes of CF$_4$ and SF$_6$ the factor $Q_i/Q$ is substantially less than 1.

Substituting (\ref{so76}) into (\ref{so66}), we find that at the temperature $T$, the mean power radiated by the $i$th mode of the molecule is
\begin{eqnarray}
\langle P_i \rangle  
&=&\frac{2\omega_i^4M_i^2 d_i }{3c^3(e^{\hbar \omega_i/k_{\rm B}T}-1)}\left(\frac{Q_i}{Q}\right)
\label{so78}
\end{eqnarray}
The power of Planck radiation at temperature $T$ absorbed by a molecule at the same temperature, is given by the sum over line intensities, $\Pi^{\{i\}}(T,T)$ of (\ref{ot6}).  In thermal equilibrium, this absorbed power must equal the spontaneously power emitted by the molecule.  If we equate (\ref{so78}), the power emitted according to the harmonic oscillator model, to (\ref{ot6}), the power absorbed from thermal radiation, we find that the transition dipole moment of the vibrating molecule must be given by
\begin{equation}
|M_i|=\sqrt{\frac{3c^3\Pi^{\{i\}}(T,T)}{2\langle n_i\rangle \omega_i^4}}.
\label{so80}
\end{equation}

In Table 5, we show values of the transition dipole moments $M_i$ given by (\ref{so80}) as well as the spontaneous emission rate

\begin{equation}
\Gamma_{n_i=1} = {{\Pi^{\{i\}}(T,T)}\over {\hbar \omega_i \langle n_i\rangle}}
\label{so81}
\end{equation}

\noindent implied by (\ref{soo56}).  The radiated powers $\Pi^{\{i\}}(T,T)$ of (\ref{ot6}) at $T=300$ K of the prominent vibrational modes evident in the spectra shown in Fig. 3 are of order $10^{-21}$ W.  The transition electric dipole moments are of order one tenth of a Debye unit (1 D $= 10^{-18}$ statC cm).  The modes of CF$_4$ at 1283 cm$^{-1}$ and SF$_6$ at 948 cm$^{-1}$ have a somewhat higher power which likely arises from the presence of fluorine which has the largest electronegativity of any atom and creates the relatively large transition moment.	 There is considerably more variation of the mean number of thermally excited vibrational quanta $\langle n_i\rangle$ and the spontaneous decay rates $\Gamma_{n_i=1}$.

The dependence of the power radiated per molecule for the range of temperatures occurring in the atmosphere is shown in Fig. 9.  From equations (\ref{so76}) and (\ref{so78}), we see the power per molecule is proportional to the mean number of vibrational excitation quanta $\langle n_i \rangle$ which increases rapidly with temperature.    

\subsection{Frequency-integrated cross sections}
For a non rotating harmonic oscillator model, we can write the Planck radiation intensity $\tilde B(\nu_{ul},T)=\tilde B(\nu_i,T)$, take it outside the sum and use (\ref{b12}) and (\ref{ot6}), with $\Pi(T,T)= \langle P_i\rangle$, to write the line intensity sum as

\begin{eqnarray}
\sum_{ul} S_{ul}^{\{i\}} &=& {{\langle P_i \rangle}\over {4 \pi \tilde B(\nu_i,T)}}\nonumber\\
&=&{{4 \pi^3 \nu_i M_i^2 d_i}\over {3 h c}} \Big({{Q_i}\over{Q}}\Big).
\label{so82}
\end{eqnarray}

\noindent The line intensity sum of (\ref{so82}) includes all line by line transition frequencies, $\nu_{ul}$, for the molecular absorption band interval $\nu_1\le \nu_{ul}\le \nu_2$, centered on the mode frequency $\nu_i$.  According to (\ref{lbl16}) to (\ref{lbl20}), this sum is the same as the frequency-integrated absorption cross section of the band,

\begin{equation}
\int_{\nu_1}^{\nu_2} \sigma (\nu,T) d\nu= \sum_{ul} S_{ul}^{\{i\}}.
\label{so83}
\end{equation}

\begin{figure}[t]
\includegraphics[height=90mm,width=0.9 \columnwidth]{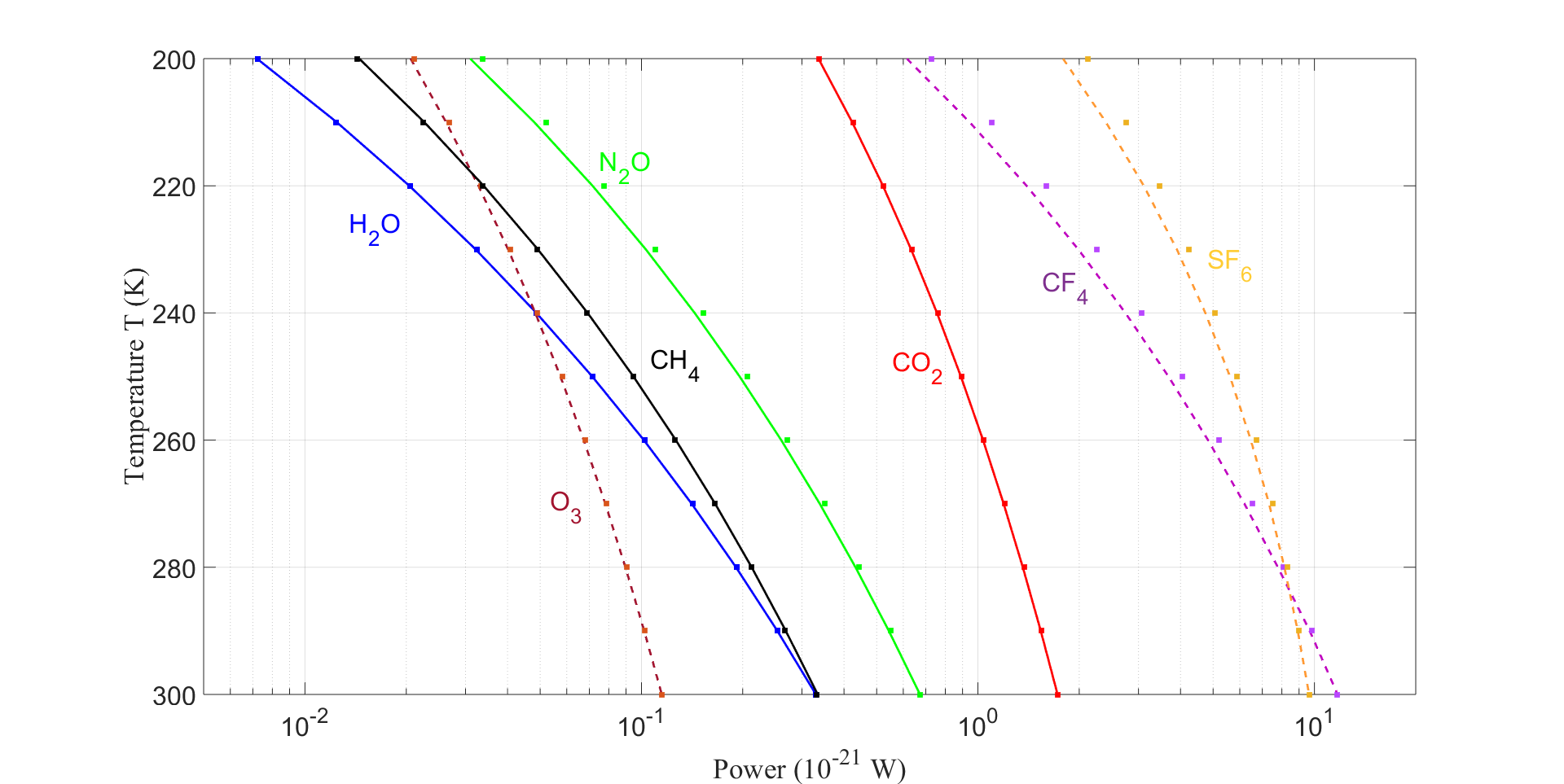}
\caption{Power radiated per molecule at a temperature $T$.   The solid lines represent the power per molecule $\Pi^{\{i\}}(T,T)$ found according to the line intensity sum of (\ref{ot6}) while the dots are the mean powers radiated by a harmonic oscillator as given by (\ref{so78}). The results are for the following modes with the frequencies in units of cm$^{-1}$: H$_2$O (1595), CO$_2$ (667), O$_3$ (699), N$_2$O (1285), CH$_4$ (1311), CF$_4$ (1283) and SF$_6$ (948). 
\label{ho}}
\end{figure}

\begin{figure}[t]
\includegraphics[height=90mm,width=0.9 \columnwidth]{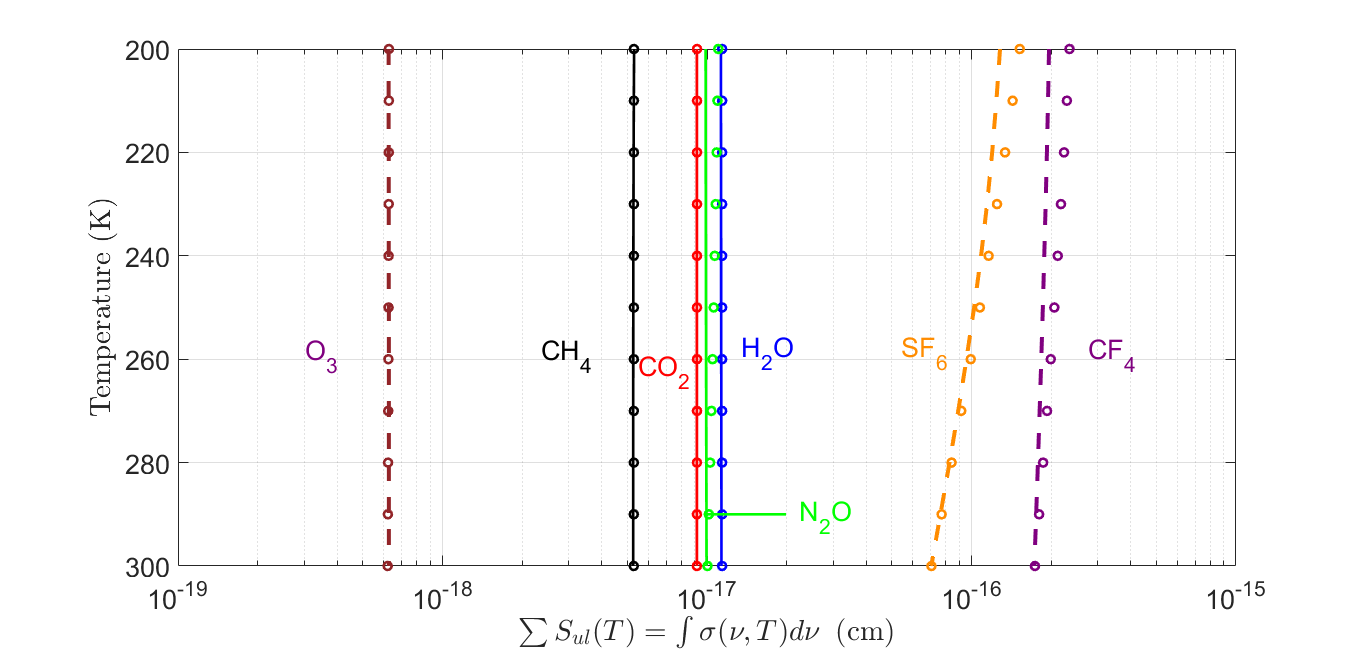}
\caption{Temperature dependence of line strength sum, $\sum_{ul} S_{ul} (T)$, or frequency integral of the cross
sections, $\int \sigma(\nu,T) d\nu$, from the right side of (\ref{so82}).  The results correspond to the following modes with the frequencies in units of cm$^{-1}$: H$_2$O (1595), CO$_2$ (667), O$_3$ (699), N$_2$O (1285), CH$_4$ (1311), CF$_4$ (1283) and SF$_6$ (948). 
\label{h1}}
\end{figure}

 Fig. 10 shows a plot of the frequency-integrated cross section versus temperature.   Equations (\ref{so82}) and (\ref{so83}) show the cross sectional temperature dependence is determined by the partition function $Q_i$ for the vibrational mode at frequency $\nu_i$ given by (\ref{so64}), divided by the total partition function $Q$ given by (\ref{so62}).  The individual partition functions, $Q_i$, increase with temperature causing $Q$ to have an even greater temperature increase.  Hence, the ratio $Q_i / Q$ decreases as the temperature increases.  The exact temperature dependence of the frequency-integrated cross section is a function of the vibrational frequencies.  For a theoretical molecule having only a single vibrational frequency, $Q_i / Q$ is unity and the frequency-integrated cross section is constant.  The temperature dependence of the frequency-integrated cross section is most noticeable for a molecule having a large number of vibrational frequencies.  This is most evident in Fig. 10 for SF$_6$ where the frequency-integrated cross section increases by nearly a factor of 2 as temperature decreases from 300 to 200 K. 

\begin{table}
	\begin{center}
	\begin{tabular}{|c c c c c c c c|}
	\hline
	Molecule & $\nu_i$  & $d_i$ &$\langle n_i\rangle$ &$\nu_1 - \nu_2$ &$\Pi^{\{i\}}(T,T)$ & $|M_i|$ & $\Gamma_{n_i=1}$ \\ [0.5ex]
	&(cm$^{-1}$) & & &(cm$^{-1}$) &(10$^{-21}$ W) & (D) &(s$^{-1}$)\\			
	\hline\hline
H$_2$O &1595 &1  &$4.75\times 10^{-4}$ &1000-2500 &0.330 &0.185 &21.9\\
	       &3652 &1  &$2.47\times 10^{-8}$ &2500-3700 &$6.40 \times 10^{-5}$ &0.068 &35.7\\
	       &3756 &1  &$1.50 \times 10^{-8}$ &3700-5000 &$5.51 \times 10^{-5}$ &0.077 &49.2\\
\hline
CO$_2$ & 667 &2  &$8.50\times 10^{-2}$ &500-850 &1.73  &0.182 &1.54\\
            &1388 &1 &$1.18 \times 10^{-3}$ & & & &\\
            &2349 &1 &$1.18\times 10^{-5}$ &1500-2500 &0.253 &0.476 &461  \\
\hline
O$_3$  & 699  &1 &$3.58 \times 10^{-2}$ &500-900 &0.115 &0.066 &0.231  \\
	     &1042 &1 &$6.53 \times 10^{-3}$ &900-1075 &1.70& 0.266 &12.6\\
           &1110 &1 &$4.70 \times 10^{-3}$ &1075-1300 &$3.23 \times 10^{-2}$ & 0.038 &0.312\\
\hline
N$_2$O & 588 &2 &$1.27\times 10^{-1}$ &500-800 &0.224 &0.069 &0.152 \\
             &1285&1 &$1.87 \times 10^{-3}$ &800-1500 &0.673 &0.206 &14.1  \\
             &2224&1 &$2.06\times 10^{-5}$ &1500-3000 &0.221 & 0.376 &243  \\
\hline
CH$_4$ &1311 &3 &$5.58\times 10^{-3}$ &1000-1900 &0.332  & 0.080 &2.28 \\ 
            &1533 &2 &$1.28 \times 10^{-3}$ & & & &\\
            &2916 &1 &$8.38\times 10^{-7}$ & & & &\\
            &3019 &3 &$1.53 \times 10^{-6}$ &1900-4000 &$2.52\times 10^{-3}$ &0.080 &27.4\\
\hline
CF$_4$ & 439 &2 & $2.35 \times 10^{-1}$ & & & &\\ 
            & 631 &3 & $1.16\times 10^{-1}$ &583-681 &0.229  & 0.063 & 0.158 \\ 
            & 922 &1 &$8.04 \times 10^{-3}$ & & & &\\
 	     &1283 &3 & $4.20\times 10^{-3}$ &895-1515 &11.7  & 0.575 & 109\\ 
\hline
SF$_6$ & 351 &3 &$3.89 \times 10^{-1}$ & & & &\\
            & 525 &3 &$1.04 \times 10^{-1}$ & & & &\\
            & 615 &3 & $5.98\times 10^{-2}$ &592-637 &1.14  & 0.207 & 1.56 \\ 
            & 643 &2 &$3.24 \times 10^{-2}$ & & & &\\
            & 775 &1 &$7.84 \times 10^{-3}$ & & & &\\
            & 948 &3 & $1.02\times 10^{-2}$ &932-964 &9.64  & 0.613 &50.2 \\ [1ex]			
\hline
	\end{tabular}
	\end{center}
\caption{Mode frequencies $\nu_i = \omega_i/(2\pi c)$, spatial degeneracies $d_i$ of (\ref{so2}) and mean number of vibrational excitation quanta $\langle n_i\rangle$ from (\ref{so76}).  For the infrared active modes, we found the radiated powers $\Pi^{\{i\}}(T,T)$ at $T=300$ K from (\ref{ot6}), for lines with frequencies $\nu_{ul}$ between the lower bound $\nu_1$ and the upper bound $\nu_2$.  Transition moments $\vert M_i\vert$ were obtained using (\ref{so80}) and the spontaneous decay rates $\Gamma_{n_i=1}$ of (\ref{so81}) for an excited molecule with one vibrational quantum.  
\label{tdm}}
\end{table}
	
\section{Conclusions}

This work examined the transmission of infrared radiation through a cloud-free atmosphere from the Earth's surface to outer space.  A line by line calculation used over 1.5 million lines of the five most important naturally occurring greenhouse gases, H$_2$O, CO$_2$, O$_3$, N$_2$O and CH$_4$ as well as CF$_4$ and SF$_6$.  This included considerably more weaker rovibrational line strengths, for H$_2$O as small as $10^{-27}$ cm, than other studies.  The calculation of forcings took into account the observed altitudinal concentrations of the various gases.

The most striking fact about radiation transfer in Earth's atmosphere is summarized by Fig. \ref{CO2}.  Doubling the current concentrations of the greenhouse gases CO$_2$, N$_2$O and CH$_4$ increases the forcings by a few percent for cloud-free parts of the atmosphere.
Table \ref{int2} shows the forcings at both the top of the atmosphere and at the tropopause for all molecules except for SF$_6$ are comparable to those found by other groups.  Our result for SF$_6$ is lower than other groups that used infrared absorption cross sections.   Increasing the concentrations of either CF$_4$ or SF$_6$ by a factor of 100 yields a forcing nearly an order of magnitude smaller than that obtained by doubling CO$_2$.

Fig. \ref{DFDC} as well as Tables \ref{int} and \ref{dPr} show that at current concentrations, the forcings from the 5 naturally occurring greenhouse gases are saturated.  The saturation of CF$_4$ and SF$_6$ is small because their concentrations are orders of magnitude less than the other greenhouse gases.  The saturations of the abundant greenhouse gases H$_2$O and CO$_2$ are so extreme that their per-molecule forcing is attenuated by four orders of magnitude with respect to the optically thin values.  Saturation also suppresses the forcing power per molecule for the less abundant greenhouse gases, O$_3$, N$_2$O and CH$_4$, from their optically thin values, but far less than for H$_2$O and CO$_2$.

Table \ref{int} and Fig. \ref{Pbar} show the overlap of absorption bands of greenhouse gases causes their forcings to be only roughly additive.  One greenhouse gas interferes with, and diminishes, the forcings of all others.  But for the case of H$_2$O, CO$_2$, O$_3$, N$_2$O and CH$_4$, the self-interference of a greenhouse gas with itself, or saturation, is a much larger effect than interference between different gases.  Table \ref{dPr} shows that for optically thin conditions, the forcing power per molecule is about a few times $10^{-22}$ W per molecule for the five naturally occurring greenhouse gases and of order $10^{-21}$ W per molecule for CF$_4$ and SF$_6$.  

An important consistency test of our work is that for optically thin conditions, the forcing power per molecule can be computed directly from the line strengths.  This was checked to agree with that found by dividing the radiative forcing by the column density.  The forcing power per molecule agreed well with the results obtained using a simple harmonic oscillator model for the various vibrational modes for all seven molecules considered in this study.  The frequency-integrated cross sections were found to be relatively independent of temperature for the five naturally occurring greenhouse molecules but not for CF$_4$ nor SF$_6$.  For SF$_6$, it varies by about a factor of 2 over the temperature range found in the atmosphere which may explain the different radiative forcings found by various groups.  In conclusion, this work is useful for examining the dependence of radiative forcing on changing greenhouse gas concentrations.

\section*{Acknowledgements}
We are grateful for constructive suggestions by many colleagues. Special thanks are due to M. Chipperfield for providing data of the SF$_6$ altitudinal profile and to I. Gordon who helped access the HITRAN data base.  The Canadian Natural Science and Engineering Research  Council provided financial support of one of us.


\begin{thebibliography}{99}
\bibitem{IPCC} G. Myhre et al, {\it Anthropogenic and Natural Radiative Forcing}, {\it Climate Change 2013:  The Physical Science Basis.  Contribution of Working Group I to the Fifth Assessment Report of the Intergovernmental Panel on Climate Change}. Cambridge University Press, Cambridge, United Kingdom (2013).

\bibitem{Hodnebrog} O. Hodnebrog et al, {\it Updated Global Warming Potentials and Radiative Efficiencies of Halocarbons and Other Weak Atmospheric Absorbers"}, Rev. Geophys. {\bf 58}, 10.1029 (2019).

\bibitem{MaunaLoa} Global Greenhouse Gas Reference Network, NOAA Earth System Research Laboratory/Global Monitoring Division, www.esrl.noaa.gov/gmc/ccgg, (2019).

\bibitem{Worton} D. R. Worton et al, {\it Atmospheric Trends and Radiative Forcings of CF$_4$ and C$_2$F$_6$ Inferred from Firn Air}, Environ. Sci. Tech. {\bf 41}, 2184 (2007).

\bibitem{Muhle} J. M\"uhle et al, {\it Perfluorocarbons in the global atmosphere:  tetrafluoromethane, hexafluoroethane and octafluoropropane}, Atmos. Chem. Phys. {\bf 10}, 5145 (2010).

\bibitem{Kovacs} T. Kovacs et al, {\it Determination of the atmospheric lifetime and global warming potential of sulfur hexafluoride using a three-dimensional model}, Atmos. Chem. Phys. {\bf 17}, 883 (2017).

\bibitem{Etminan2016} M. Etminan, G. Myhre, E. J. Highwood and K. P. Shine, {\it Radiative Forcing of Carbon Dioxide, Methane and Nitrous Oxide: A Significant Revision of the Methane Radiative Forcing}, Geophys. Res. Lett. {\bf 43}, 12614 (2016).

\bibitem{Edwards1992} D. P. Edwards, {\it A General Line-by-Line Atmospheric Transmittance and Radiance Model}, NCAR Technical Note, NCAR/TN-367+STR (1992).

\bibitem{Collins2006} W. D. Collins {\it et al.} {\it Radiative Forcing by Well Mixed Greenhouse Gases:  Estimates from Climate Models in the Intergovernmental Panel on Climate Change (IPCC) Fourth Assessment Report (AR4)}, J. Geophys. Res. {\bf 111}, D14317 (2006).

\bibitem{Schreier2014} F. Schreier, S. G. Garcia, P. Hedelt, M. Hess, J. Mendrok, M. Vasquez and J. Xu, {\it GARLIC -- A General Purpose Atmospheric Radiation Transfer Line-by-Line Infrared-Microwave Code: Implementation and Evaluation}, JQSRT {\bf 137}, 29 (2014).

\bibitem{vW2020} W. A. van Wijngaarden and W. Happer, {\it Dependence of Earth's Thermal Radiation on Five Most Abundant Greenhouse Gases}, Atmos. \& Oceanic Phys. arXiv: 2006.03098 (2020).

\bibitem{HITRAN} I. E. Gordon, L. S. Rothman et al., {\it The HITRAN2016 Molecular Spectroscopic Database}, JQSRT {\bf 203}, 3-69 (2017).

\bibitem{CNRS} Virtual Atomic and Molecular Data Centre, VAMDC Consortium, https://vamdc.icb.cnrs.fr (2020).

\bibitem{Temp} {\it The U.S. Standard Atmosphere}, NASA Report TM-X-74335 (1976).

\bibitem{Anderson} G. P. Anderson, S. A. Clough, F. X. Kneizys, J. H. Chetwynd and E. P. Shettle, {\it AFGL Atmospheric Constituent Profiles (0-120 km)}, AFGL-TR-86-0110  Air Force Geophysics Laboratory, Hanscom Air Force Base, Massachusetts, (1986).

\bibitem{Trudinger} C. M. Trudinger et al, {\it Atmospheric abundance and global emissions of perfluorocarbons CF$_4$, C$_2$F$_6$ and C$_3$F$_8$ since 1800 inferred from ice core, firn, air archive and in situ measurements}, Atmos. Chem. Phys., {\bf 16}, 11733 (2016).

\bibitem{Schwarzschild1906} K. Schwarzschild, {\it \"Uber Diffusion und Absorption in der Sonnenatmosph\"are}, Nachr. K. Gesell. Wiss. Math.-Phys. Klasse {\bf 195}, 41 (1906).

\bibitem{Buglia} J. J. Buglia, {\it Introduction to the Theory of Atmospheric Radiative Transfer}, NASA Publication 1156 (1986).

\bibitem{Wilber} A. C. Wilber, D. P. Kratz and S. K. Gupta, {\it Surface Emissivity Maps for Use in Satellite Retrievals of Longwave Radiation}, NASA/TP-1999-209362 (1999).

\bibitem{Yoshikawa} K. K. Yoshikawa, {\it An Iterative Solution of an Integral Equation for Radiative Transfer using Variational Technique}, NASA Technical Report TN-D-7292, A-4774 (1973).

\bibitem{Zhong2013} W. Zhong and J. D. Haigh, {\it The Greenhouse Effect and Carbon Dioxide}, Weather {\bf 68}, 100 (2013).

\bibitem{Hurley} M. D. Hurley, T. J. Wallington, G. A. Buchanan, L. K. Gohar, G. Marston and K. P. Shine, {\it IR spectrum and radiative forcing of CF$_4$ revisited}, J. Geophys. Res. {\bf 10}, D02102 (2005).

\bibitem{HarrisonCF4} J. J. Harrison {\it New infrared absorption cross sections for the infrared limb sounding of carbon tetrafluoride (CF$_4$)}, JQSRT {\bf 260}, 107432 (2021).  

\bibitem{HarrisonSF6} J. J. Harrison, {\it New infrared absorption cross sections for the infrared limb sounding of sulfur hexafluoride (SF$_6$)}, JQSRT {\bf  254}, 107202 (2020).

\bibitem{Landau} L. Landau and E. Lifschitz, {\it Quantum Mechanics (Nonrelativsitic Theory)}, Addison Wesley, Reading, Mass. (1965).


\end{thebibliography}
\end{document}